\documentclass[trackchanges]{aastex631}

\usepackage{graphicx}
\graphicspath{ {./images/} }
\usepackage{textcomp}
\usepackage{gensymb}
\usepackage{amsmath}
\usepackage{hyperref}
\usepackage{comment}
\usepackage{wrapfig}
\usepackage{upgreek}
\usepackage{longtable}
\usepackage{lipsum}
\usepackage[version=4]{mhchem}

\begin{document}

\title{Whole-disk Photometric and Colorimetric Phase Curves of Earth with the \textit{Clementine} Lunar Orbiter}

\author[0009-0008-1520-4272]{Chase Cooper}
\affiliation{Lunar and Planetary Laboratory, University of Arizona, Tucson, AZ 85721, USA}
\affiliation{Department of Astronomy, University of Arizona, Tucson, AZ 85721, USA}
\affiliation{Habitability, Atmospheres, and Biosignatures Laboratory, University of Arizona, Tucson, AZ 85721, USA}

\author[0000-0002-3196-414X]{Tyler D. Robinson}
\affiliation{Lunar and Planetary Laboratory, University of Arizona, Tucson, AZ 85721, USA}
\affiliation{Habitability, Atmospheres, and Biosignatures Laboratory, University of Arizona, Tucson, AZ 85721, USA}
\affiliation{NASA Nexus for Exoplanet System Science Virtual Planetary Laboratory, University of Washington, Box 351580, Seattle, WA 98195, USA}

\author[0000-0002-7755-3530]{Jason W. Barnes}
\affiliation{Department of Physics; University of Idaho; Moscow, ID 83844, USA}

\author[0000-0002-4321-4581]{L.~C. Mayorga}
\affiliation{Johns Hopkins Applied Physics Laboratory, Laurel, MD, 20723, USA}

\author[0009-0004-9747-5623]{Lily Robinthal}
\affiliation{Lunar and Planetary Laboratory, University of Arizona, Tucson, AZ 85721, USA}
\affiliation{Habitability, Atmospheres, and Biosignatures Laboratory, University of Arizona, Tucson, AZ 85721, USA}

% Rewrite!!!!!
\begin{abstract}
    The phase curve on a planetary object can reveal key details about its atmosphere and surface, including the presence of surface liquid oceans. Phase curves for Earth can inform the search for habitable surface environments on terrestrial exoplanets, but such phase curves are quite rare. The \textit{Clementine} lunar orbiter made 3,744 observations of the Earth between February and May of 1994. These observations spanned phase angles between $3.4\degree$ and $139.1\degree$. From these data, we produce phase curves of Earth in 5 distinct bands with wavelengths between 400\,nm and 1000\,nm. We extrapolate fits to the phase curves to derive the geometric albedo of Earth in each filter. We examine the Earth's color as a function of phase angle and find the observed values are best fitted by models that include contributions from ocean glint. We also find a number of observations where contributions from glint are minimal but the disk-integrated color is in line with the glinting model predictions; we propose that this can be explained by the presence of optically clouds on the disk. These findings stress the importance of high-phase Earth observations for high-fidelity Earth models and for future efforts to characterize terrestrial exoplanets via direct imaging.
    
    % Earth is the quintessential habitable planet, and perhaps its key feature (and a strong habitability indicator) is its global ocean. Oceans are predicted to be detectable on terrestrial exoplanets via the specular reflection of starlight off their surface, termed ``ocean glint.'' We employ a pipeline that reduces whole-disk observations of Earth, acquired by the \textit{Clementine} lunar orbiter, into point source observations. From these data, we present novel phase curves of Earth at 5 distinct bands that showcase the planet's non-Lambertian scattering behavior. When comparing these observations to a bespoke 3D Earth model, we find that observations favor a model with glint over one without glint. The data are analyzed to look for colorimetric evidence of glint at high phase angles, but too few data at sufficiently high phase angles are present to conclusively determine the presence or absence of glint. These findings stress the importance of more observations of Earth at extreme crescent phases to adequately describe the contribution of ocean glint to planetary phase angles. Such data would inform the development of future telescopes, such as the Habitable Worlds Observatory, which will search for potentially habitable planets beyond the Solar System.
\end{abstract}


\section{Introduction}\label{sec:intro}

Models of Earth predict that its phase curves should contain a wealth of information about the planet's surface and atmosphere. Different scattering processes, such as by clouds, aerosols, or oceans, manifest in phase curves in different ways, and can thus be disentangled from one another. Indeed, phase curves of various solar system objects have illuminated aspects of their atmospheres and environments. Phase curves of Venus, both photometric \citep{Arking_1968} and polarimetric \citep{Coffeen_1969}, were used to identify micron-sized sulfuric acid droplets within Venusian clouds \citep{Young_1973}. Recent analyses of multi-phase \textit{Cassini} observations of Titan showed that hazes in its atmosphere cause such strong forward scattering that Titan appears brighter near new phase than it does at full phase \citep{GarciaMunoz_2017,Cooper_2025}. Phase curves of Saturn's rings have been used to detail the size distributions and compositions of their constituent materials \citep[e.g.][]{Franklin_1965,Showalter_1992,Poulet_2002}. Moving beyond the solar system, thermal emission phase curves have been used to infer the thermal and atmospheric properties of an increasing amount of exoplanets \citep[e.g.][]{Knutson_2012,Wong_2020,Mikal-Evans_2023,Kempton_2023,Dang_2025}.

The specular reflection of light by an ocean, or ocean glint, is a scattering process that could manifest in planetary phase curves \citep{McCullough_2006,Williams_2008}. Ocean glint produces the greatest signal at high phase angles where water is most reflective and the necessary geometry to specularly reflect is more often achieved by surface waves \citep{Cox_1954}. This effect is strong enough that high-fidelity Earth models predict the broadband visible reflectance of Earth is up to 100\% higher at extreme crescent phases than it would be without glint \citep{Robinson_2010,Zugger_2010}. 

The prospect of detecting glint in planetary phase curves is important to the study of potentially habitable exoplanets \citep{Robinson_2018}. Glint in an exoplanetary phase curve may be a viable method of detecting surface oceans on another world, a key indicator of potential habitability. Future observatories, including the Habitable Worlds Observatory \citep[HWO;][]{feinbergetal2024}, will enable the construction of phase curves for nearby rocky planets for the first time. Ocean glint from nearby ocean worlds in (near-)edge-on orbits will be detectable by HWO \citep{Vaughn_2023}.

% Much work has gone into retrieving phase curves of Earth, using both Earthshine, sunlight reflected by the Earth towards the Moon which is then reflected back to the Earth , and occasional whole-disk observations of the Earth from interplanetary spacecraft, . The former group are historically more abundant and span a greater range of phase angles, while the latter group are sparse but provide direct measurements of the disk-integrated reflectivity of Earth. These data together were recently used by \citet{Robinson_2025} to derive the broadband geometric albedo and phase curve of Earth, as well as to investigate how different scattering processes, including glint, may contribute thereto. However, measurements of Earth's reflectivity at extreme crescent phases ($\alpha\ge130\degree$), where the contribution from glint is theorized to be greatest, are very limited. 

Whole-disk photometric observations of the Earth are limited in number and are difficult to obtain. Observations from spacecraft, such as from \textit{EPOXI} \citep{Livengood_2011}, \textit{LCROSS} \citep{Robinson_2014}, and \textit{Galileo} \citep{Strauss_2024}, offer direct measurements of the Earths reflectivity but with a limited range of phase angles. Ground-based measurements of Earthshine \citep{Danjon_1936,Goode_2001,Palle_2003,Qiu_2003} are more numerous and cover a greater range of phase angles, but are typically limited to broadband visible light, though Earthshine observations in the NIR \citep{Turnbull_2006} and visible light polarization \citep{Roccetti_2025c} exist as well. Measurements from either source at extreme crescent phases ($>130\degree$), where models predict the greatest contribution from glint, are rare.

An unexplored source of whole-disk Earth observations comes from the \textit{Clementine} lunar orbiter, which launched in 1993 to image the lunar surface in detail before departing for an encounter with the near-Earth asteroid 1620 Geographos \citep{Nozette_1994}. However, a computer issue aboard \textit{Clementine} prevented it from carrying out observations during the second half of its mission. Onboard \textit{Clementine} were a number of scientific instruments, including the Ultraviolet/Visible Camera (``UV/Vis'') \citep{Kordas_1995}. The UV/Vis imager was equipped with five filters that spanned visible and near-infrared wavelengths between 415\,nm and 1000\,nm, labeled filters A through E, as well as a broadband visible filter, filter F. From February to May 1994, \textit{Clementine} repeatedly observed the Earth, producing a rare set of whole-disk, photometrically diverse images of our planet over a wide range of phase angles.

The aim of this work is to present novel, disk-averaged \textit{Clementine} observations of Earth, including at extreme phase angles where ocean glint is relevant. Additionally, we attempt to measure the contribution of glint to the reflected-light phase curves of Earth and assess the feasibility of detecting glint in the phase curves of a directly imaged, ocean-covered planet. Section~\ref{sec:methods} covers the methodology used to acquire, calibrate, and spatially reduce the \textit{Clementine} images. In Section~\ref{sec:results}, we present the derived phase curves of Earth from the \textit{Clementine} data. Section~\ref{sec:discussion} examines the photometric and spectral signatures of glint in these products, and whether or not these signatures are within reach of next-generation observatories like HWO. To conclude, Section~\ref{sec:conclusions} summarizes the findings of this works and outlines its implications on future observations.

\section{Methods}\label{sec:methods}

\begin{figure*}[t]
    \centering
    \includegraphics[width=\linewidth]{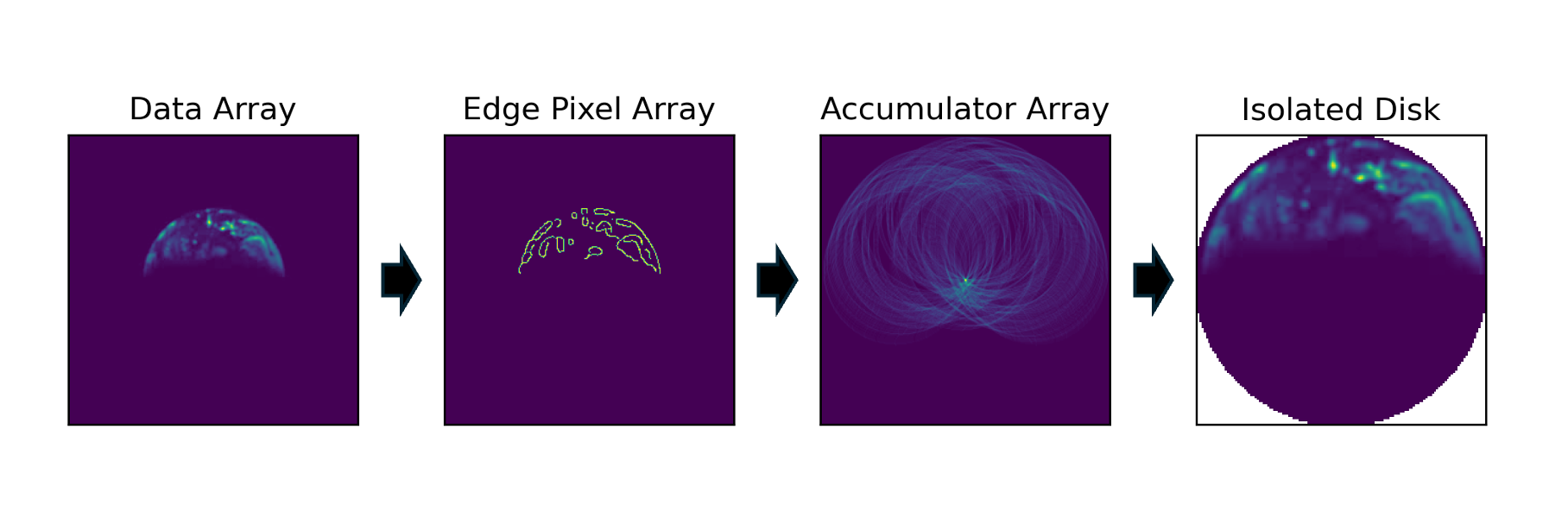}
    \caption{A visual example of how our pipeline isolates the disk of Earth from a \textit{Clementine} cube. Far left: the original cube. Middle left: the edge pixel array determined using a Canny edge detection algorithm. Middle right: the accumulator array; the brightest pixel in the array is assumed to be the center of the disk of Earth. Far right: the isolated disk of Earth.}
    \label{fig:diskdetect}
\end{figure*}

In the following section, we describe the pipeline with which we turn raw, spatially resolved observations (``cubes'') of Earth to disk-averaged values of Earth's reflectivity. Section~\ref{subsec:methacq} explains how the data were acquired and calibrated, Section~\ref{subsec:methdisk} describes how our pipeline automatically detects the disk of Earth in each image, and Section~\ref{subsec:methcorr} goes over the method used to correct saturated pixels in the cubes. Section~\ref{subsec:methcalc} describes how values across the disk of Earth are converted to a disk-averaged value. Finally, Section~\ref{subsec:methobs} the spatial and temporal distribution of the cubes.

\subsection{Data Acquisition and Calibration}\label{subsec:methacq}

\begin{deluxetable}{cc}[b]
    \tablecaption{Filters used to remove invalid \textit{Clementine} cubes \label{table:criteria}}
    \tablenum{1}
    \tablehead{\colhead{Filter Criterion} & \colhead{\# of Cubes Removed}}
    \startdata
    Filter F & 590 \\
    Failed calibration & 84 \\
    Erroneous pixels $> 5\%$ & 226 \\
    Saturated illuminated pixels $> 5\%$ & 59 \\
    Manual sort & 55\\
    \hline 
    Total invalid cubes & 1,014 \\
    Valid cubes & 2,730 \\
    \enddata
\end{deluxetable}

\begin{deluxetable*}{cccccc}[t]
    \tablecaption{UV/Vis filter properties and cube counts. The corrective factors are sourced from \citet{Ohtake2010}.
    \label{table:filter}}
    \tablenum{2}
    \tablehead{\colhead{Filter ID} & \colhead{$\lambda_{\rm eff}$ [nm]} & \colhead{Bandwidth [nm]} & \colhead{Corrective Factor} & \colhead{\# of cubes} & \colhead{\# of usable cubes}}
    \startdata
    A & 412.9 & $\pm20$ & 0.637 & 630 & 541 \\
    B & 750.5 & $\pm5$ & 0.673 & 632 & 541 \\
    C & 897 & $\pm10$ & 0.716 & 627 & 544 \\
    D & 951.4 & $\pm15$ & 0.729 & 632 & 559 \\
    E & 1000.2 & $\pm15$ & 0.738 & 632 & 545 \\
    F & 650 & $\pm250$ & --- & 590 & 0 \\
    \hline
    Total & --- & --- & --- & 3,744 & 2,730 \\
    \enddata
    
\end{deluxetable*}

The data used in this work were acquired using the Planetary Data System's Image Atlas. In gathering our \textit{Clementine} dataset, we downloaded all data products produced by UV/Vis whose explicit target was Earth, for a total of 3,744 cubes. Cubes were then passed through a calibration process using the United States Geological Survey's \texttt{ISIS3} software \citep{Rodriguez2024}. Each cube is first reformatted to an \texttt{ISIS3}-compatible format that enables further calibration. Information on the spacecraft's location, pointing, and target information is added using \texttt{SPICE}, a toolkit designed for providing ancillary instrument data needed for interpreting data from spacecraft \citep{ACTON199665,ACTON20189}. A final \texttt{ISIS3} subroutine performs radiometric corrections on the cubes and converts data numbers to reflectances. Due to a lack of calibration files for UV/Vis filter F, we were unable to use \textit{Clementine} cubes taken with this filter. An additional 84 \textit{Clementine} cubes at other filters failed calibration and were excluded from our analysis.

After calibration, we used additional criteria to remove cubes that were not of sufficient quality for this work. Cubes where more than 5\% of pixels contained erroneous data were removed, as were cubes where more than 5\% of illuminated pixels were saturated. This latter criterion preferentially removed high-phase observations in all filters, but did not restrict the range of phase angles in the sample. We then pass all cubes through a manual sort to remove cubes that only image part of Earth's disk or suffer from other apparent errors. In total, we had 2,730 \textit{Clementine} cubes for analysis. A breakdown of filter criteria and the quantity of \textit{Clementine} cubes removed by each criterion is shown in Table~\ref{table:criteria}.

\subsection{Disk Detection}\label{subsec:methdisk}

To arrive at disk-averaged values of Earth's reflectivity, it was necessary to identify Earth's disk within a given cube. We used a pipeline that automatically identifies the disk of Earth within a cube using a strategy adapted from \citet{Strauss_2024}. First, the pipeline employs a Canny edge detection algorithm \citep{Canny_1986}, which identifies structures in images by looking for bright pixel-dark pixel boundaries. In a \textit{Clementine} cube of Earth, this may be due to land-water boundaries, cloud-water boundaries, or, most critically, the boundary between Earth and deep space. Edge pixels identified by the Canny edge detection algorithm have their positions stored in a separate edge pixel array the same size as the cube.

Next, the pipeline performs a circle Hough transform on the edge pixel array \citep{xie_2002}. A circle Hough transform is a method of identifying circular structures with a known radius in an image. First, an empty ``accumulator array'' the same size as the edge pixel array is created. For each pixel in the edge pixel array, a circle with the radius of the Earth (in pixels) is overlaid on the accumulator array with its center at the location of the given edge pixel, and each pixel the circle falls on has its value incremented by 1. At the end of this process, the pixel in the accumulator array with the highest value is taken to be the center of the circular feature. All pixels within the radius of the Earth of this pixel are then identified as lying on the disk of Earth. We calculated the radius of Earth in a given cube as,
\begin{equation}
    R_{\rm pixel} = \bigg\lceil \frac{1}{\Omega}\tan^{-1}{\bigg(\frac{R}{d_{\rm sc}}\bigg)} \bigg\rceil
\end{equation}
where $\Omega=255$\,$\upmu$rad is the field of view per side of a pixel, $R = 6,371$\,km is the equatorial solid body radius of Earth, and $d_{\rm sc}$ is the distance of \textit{Clementine} from Earth at the time of cube acquisition. We added 1 to $R_{\rm pixel}$ to account for the instrument's point spread of 1.1 to 1.5 pixels \citep{Kordas_1995} and atmospheric limb brightening effects that may cause Earth to appear slightly larger than its solid-body radius; we found that increasing $R_{\rm pixel}$ by more than 1 increased the calculated reflectivity of the disk by less than 0.001. A visual summary of the disk detection process is shown in Figure~\ref{fig:diskdetect}.

\subsection{Saturated Pixel Correction} \label{subsec:methcorr}

Though we culled \textit{Clementine} cubes with excessive saturated pixels, we were still left with some cubes with minor saturation issues that needed to be corrected. The saturated pixels were correlated with high-albedo phenomena like clouds or, at high phase angles, glint. This work focuses heavily on the effect of glint in disk-averaged Earth observations, so it is important that any remaining saturation issues are handled carefully.

The correction of saturated pixels was accomplished by using two methods. In the first method, we included in our pipeline a method for producing a ``glint mask'' that can be overlaid on the \textit{Clementine} image. The glint mask is produced by using a simple glint model to predict the extent of the glint spot and approximate reflectances for pixels within the glint spot (see Appendix~\hyperref[sec:appendix_a]{A}). The glint calculation takes as a parameter the wind speed over the relevant body of water \citep{Cox_1954}. Low wind speeds produce flatter waters that form smaller, more intense glint spots while higher wind speeds produce dimmer, broader glint spots. In calculating the glint mask, the wind speed parameter is selected to have the value between 0\,km/s and 15\,km/s which minimizes the total residuals between the unsaturated \textit{Clementine} pixels and the glint spot prediction over the same pixels. This fitted glint mask was then used to replace saturated pixels that fall within the glint spot.

In a second method, we made the assumption that saturated pixels should generally be at least as bright as the brightest nonsaturated pixels. Thus, we can fill the saturated pixels with a ``flat fill'' where every saturated pixel is replaced with the highest nonsaturated pixel value. This method produced a lower bound on the actual values for pixels in the disk; when calculating disk-averaged measurements, as described in the following section, we used the larger of the two values produced using the two methods detailed here.

\subsection{Reflectivity Calculation \label{subsec:methcalc}}

\begin{figure*}
    \centering
    \includegraphics[width=\linewidth]{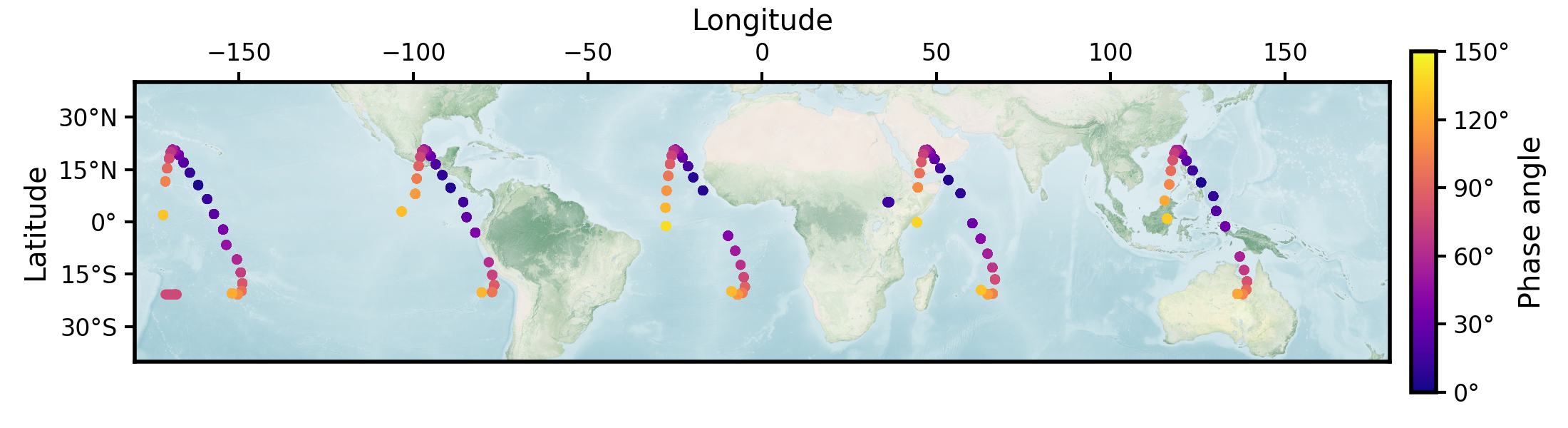}
    \caption{The distribution of (approximate) sub-spacecraft coordinates of each \textit{Clementine} cube. Points are colored depending on the phase angle of Earth at time of observation.}
    \label{fig:lonlat_dist}
\end{figure*}

The cube pixel values after calibration are measurements of $\pi\cdot I/F$, where ${I/F}$ is a measure of an object's reflectivity per steradian in units of inverse steradians. We integrate these values over the disk of Earth to get a disk-averaged measure of Earth's reflectivity, $A_{\rm g}\Phi(\alpha)$, using,
\begin{equation}
    A_{\rm g}\Phi(\alpha) = \pi\frac{\bar{I}}{F} = \frac{d_{\rm sc}^2\Omega}{\pi R^2} \sum_i (\pi\cdot I/F)_{i} \ ,
\end{equation}
where $\alpha$ is the phase angle of the observation, $\bar{I}/{F}$ is the disk-averaged reflectivity per steradian of Earth, and the sum represents the sum of reflectance values from all pixels on the disk. The resulting quantity, $A_{\rm g}\Phi(\alpha)$, is a measure of the phase-dependent reflectivity of Earth. Critically, $A_{\rm g}\Phi(\alpha)$ is the product of two quantities: the geometric albedo, $A_{\rm g}$, of the object, and the planetary phase function, $\Phi(\alpha)$. The phase function is normalized to 1 at full phase, so values of $A_{\rm g}\Phi(\alpha)$ at small phase angles approximate the geometric albedo.

Previous works \citep{Hillier1999,Shkuratov2001} have noted discrepancies between radiometric measurements of lunar features from \textit{Clementine} and from ground-based observatories. As a remedy, we applied the corrective factors derived by \citet{Ohtake2010} (see Table~\ref{table:filter}) to our disk-averaged reflectivity values. The UV/Vis instrument was calibrated using laboratory measurements of the reflectivity of lunar regolith from the Apollo 16 landing site; \citet{Ohtake2010} produce their corrective factors by taking the ratio of the reflectivities of the landing site measured with the SELENE spacecraft's Multiband Imager and Spectral Profiler and the UV/Vis laboratory measurements.

In studying the phase-dependent reflectivity of an object, we found it helpful to convert $A_{\rm g}\Phi(\alpha)$ to apparent albedo, $A_{\rm app}(\alpha)$. The apparent albedo of an object is the albedo of a Lambert sphere necessary to reproduce the observed reflectivity of that object, and it is defined as,
\begin{equation}\label{eq:a_app}
    A_{\rm app}(\alpha) = \frac{3}{2}\frac{A_{\rm g}\Phi(\alpha)}{\Phi_{\rm L}(\alpha)} \ ,
\end{equation}
where,
\begin{equation}
    \Phi_{\rm L}(\alpha) = \frac{\sin(\alpha)+(\pi-\alpha)\cos(\alpha)}{\pi} \ ,
\end{equation}
is the phase function of a Lambert disk \citep{Russell1916}.
The factor of 3/2 is due to the geometric albedo of a Lambert sphere being two-thirds of its spherical albedo. By dividing out the diffuse contributions to a planetary object's reflectivity, the apparent albedo is particularly sensitive to non-diffuse scattering effects due to the surface and atmospheric composition of that object. At full phase, the equality in Eq.~\ref{eq:a_app} reduces to
\begin{equation} \label{eq:a_app_full}
    A_{\rm app}(\alpha=0)=\frac32A_{\rm g} \ .
\end{equation}

\subsection{Observation Properties \label{subsec:methobs}}

\textit{Clementine} data in our dataset were largely acquired during one lunar orbit in April 1994; a small number of cubes originated from the previous orbit during March 1994. The phase angles of these cubes range from 3.4$\degree$ to 139.5$\degree$. As a lunar orbiter, the bounds on the phase angles of Earth \textit{Clementine} could access are approximately equal to those of the Moon; the $\sim5.2\degree$ inclination of the Moon relative to the ecliptic means the lowest accessible phase angle in a given orbit is not necessarily 0$\degree$. 

Cubes were collected in batches of about 15--30, in regular 5- to 6-hour intervals, typically including multiple cubes acquired with each UV/Vis filter. The acquisition of cubes in these batches means that data points often appear grouped together in later figures. The exact sub-spacecraft coordinates (SSCs) of \textit{Clementine} observations are not provided in the calibrated files, and exact \textit{Clementine} SSCs could not be retrieved using the JPL Horizons system. We instead use the coordinates of the Earth sub-lunar point for each \textit{Clementine} cube, which is provided in the calibrated cube as \texttt{SUB\_LIGHT\_SOURCE\_LONGITUDE} and \texttt{SUB\_LIGHT\_SOURCE\_LATITUDE}. At \textit{Clementine}'s greatest elongation from the Moon, the sub-lunar coordinates differ from the SSCs by less than $0.5\degree$, so these are a suitable stand-in for the true SSCs. The adopted SSCs are shown in Figure~\ref{fig:lonlat_dist}.

\section{Results} \label{sec:results}

\begin{figure*}
    \centering
    \includegraphics[width=\linewidth]{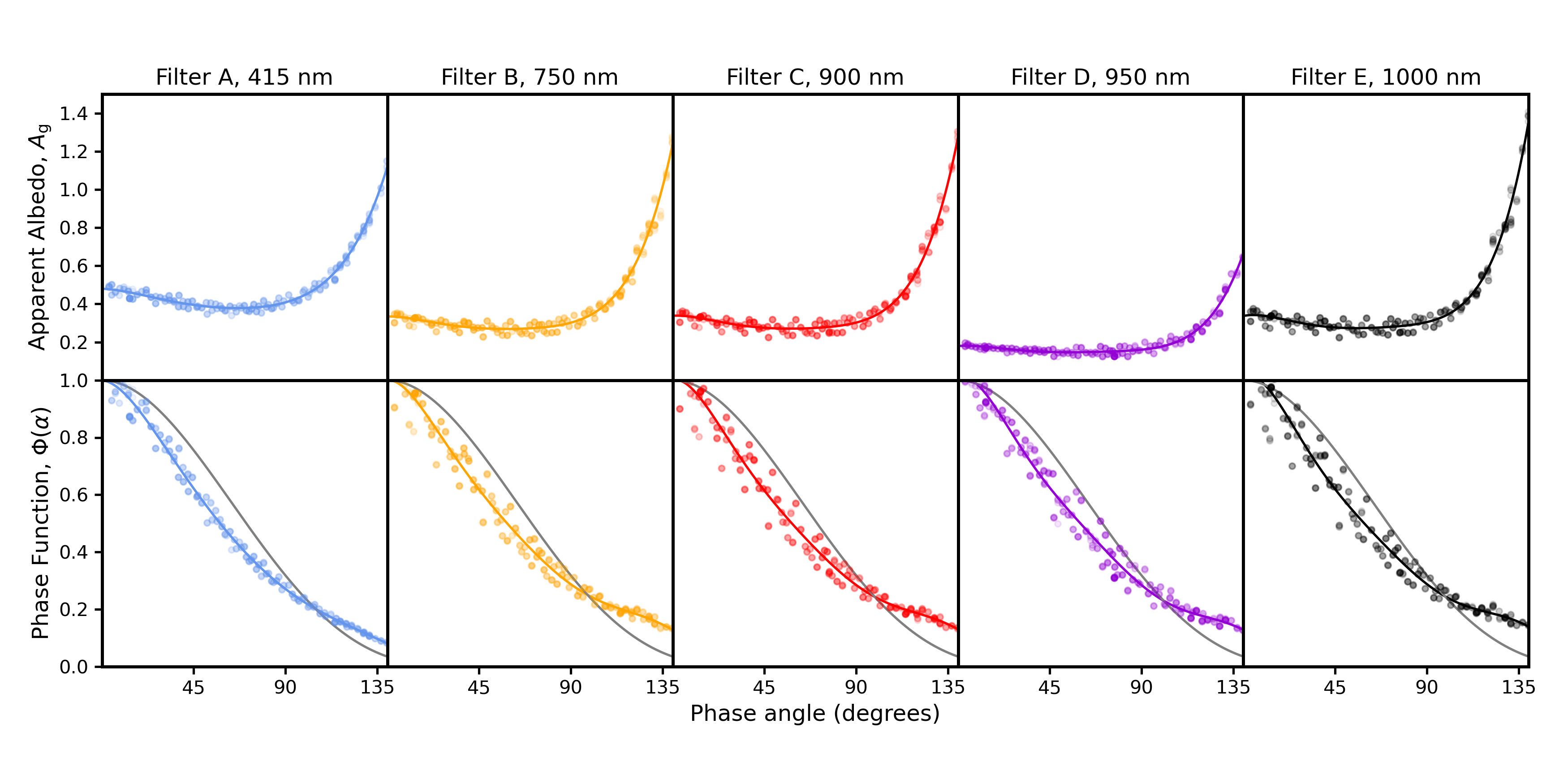}
    \caption{Apparent albedo phase curves (top row) and planetary phase functions (bottom row) of Earth at each UV/Vis filter. Fifth-degree polynomials fitted to each data subset are also shown. The Lambert phase function is shown in grey in each planetary phase function plot.}
    \label{fig:phase_curves}
\end{figure*}

Figure~\ref{fig:phase_curves} presents apparent albedo phase curves and phase functions of Earth in the five filters measured by UV/Vis. Fifth-degree polynomials fitted to the data are also shown. Each phase curve is composed of $>500$ \textit{Clementine} observations. The phase functions are the result of normalizing the data in units of $A_{\rm app}$ by the value of a fifth-degree polynomial at $\alpha=0\degree$ fitted to the data; as a result, some individual data points have $\Phi(\alpha)>1$. Uncertainties in the data are dominated by instrument calibration, which is less than 5\% \citep{Kordas_1995}.

The apparent albedo phase curves of Earth showcase distinctly non-Lambertian behavior: below $\sim100\degree$, the planet is sub-Lambertian, increasingly so up to an angle of about $60\degree$. Above $\sim100\degree$, strong wavelength-dependent forward scattering is apparent. The phase function of Earth is strictly decreasing. The deviation of points from the fit on a given graph often exceeds the calibration uncertainty. Potential causes of these phenomena are diverse and are discussed in Section~\ref{sec:discussion}. 

\begin{figure}
    \centering
    \includegraphics[width=0.5\linewidth]{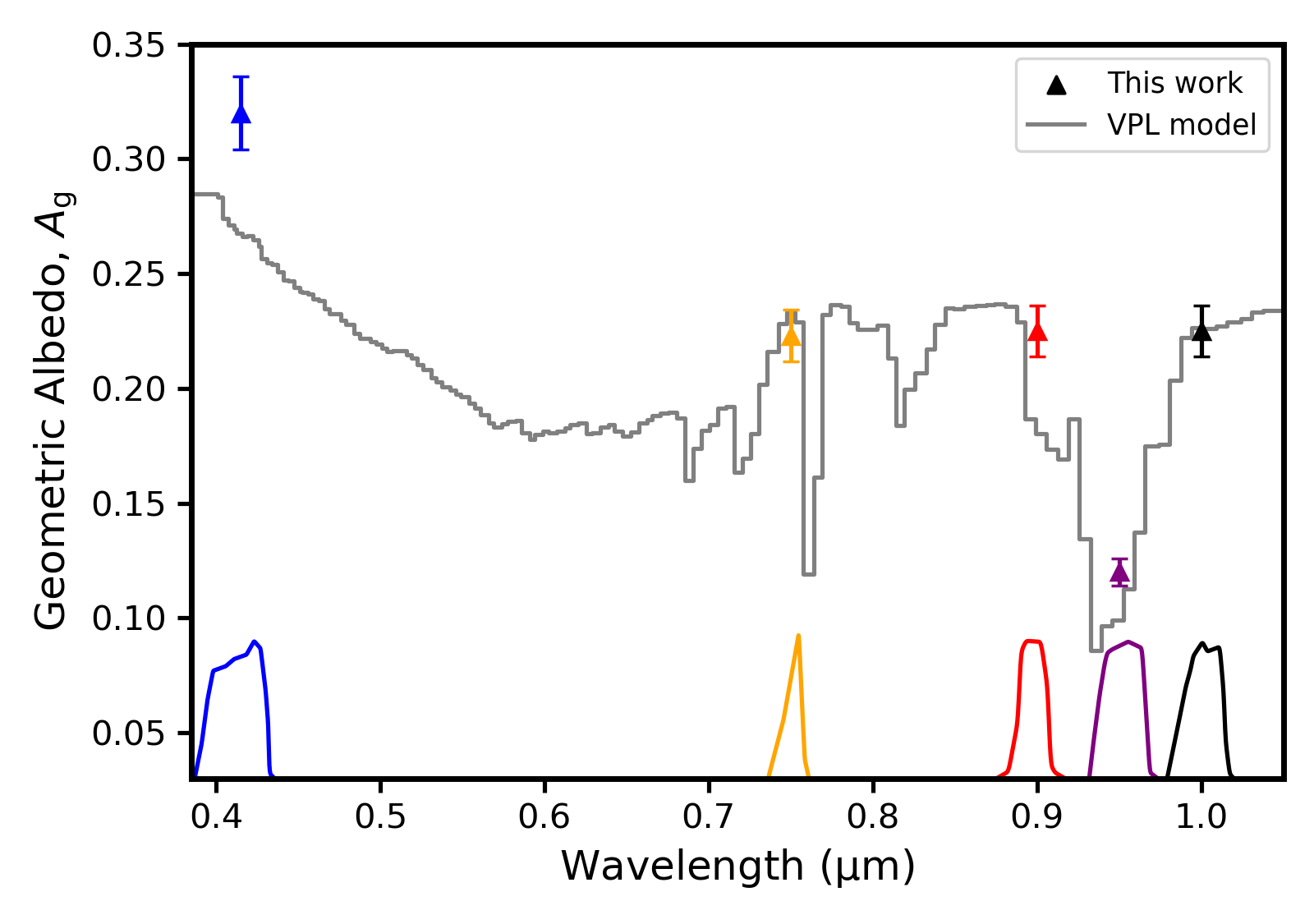}
    \caption{Derived geometric albedos of Earth within each UV/Vis filter with 5\% calibration uncertainties indicated with error bars. Values from this work are shown with the colored triangles; the modeled spectrum from the VPL Earth model \citep{Robinson_2010} is shown in grey. The filter functions of each \textit{Clementine} filter are depicted at the bottom of the figure.}
    \label{fig:geoalb}
\end{figure}

\begin{deluxetable*}{c|cccccc}
    \tablecaption{Polynomial coefficients for the best fits to Earth's apparent albedo phase curves in each filter.
    \label{table:poly}}
    \tablenum{3}
    \tablehead{\colhead{Filter ID} & \colhead{$x^5$} & \colhead{$x^4$} & \colhead{$x^3$} & \colhead{$x^2$} & \colhead{$x^1$} & \colhead{$x^0=\frac32A_{\rm g}$}}
    \startdata
    A & $8.803\cdot10^{-11}$ & $-2.258\cdot10^{-8}$ & $2.445\cdot10^{-6}$ & $-1.061\cdot10^{-4}$ & $-3.614\cdot10^{-4}$ & $0.480$ \\
    B & $1.517\cdot10^{-10}$ & $-3.759\cdot10^{-8}$ & $3.552\cdot10^{-6}$ & $-1.322\cdot10^{-4}$ & $2.092\cdot10^{-4}$ & $0.334$ \\
    C & $1.759\cdot10^{-10}$ & $-4.470\cdot10^{-8}$ & $4.285\cdot10^{-6}$ & $-1.629\cdot10^{-4}$ & $6.037\cdot10^{-4}$ & $0.338$ \\
    D & $1.166\cdot10^{-10}$ & $-3.084\cdot10^{-8}$ & $3.003\cdot10^{-6}$ & $-1.171\cdot10^{-4}$ & $8.265\cdot10^{-4}$ & $0.179$ \\
    E & $2.314\cdot10^{-10}$ & $-6.125\cdot10^{-8}$ & $6.043\cdot10^{-6}$ & $-2.418\cdot10^{-4}$ & $1.925\cdot10^{-3}$ & $0.337$ \\
    \enddata
\end{deluxetable*}

We extrapolate the polynomials fitted to Earth's apparent albedo phase curves to full phase to get approximate full-phase apparent albedos, and with that we estimate the geometric albedo of Earth at the five UV/VIS wavelengths using Eq.~\ref{eq:a_app_full}. The derived geometric albedos are shown in Figure~\ref{fig:geoalb}. Because Earth possesses a significant atmosphere and is largely covered by water, we do not consider a potential opposition effect, which is most prominent on airless, rocky bodies. This is supported by the data, which do not show any such effect down to phase angles of $3.4\degree$, and by \citet{Robinson_2025}, who found no opposition effect in Earth's broadband visible light phase curve. The inferred values for UV/VIS filters B through E are in agreement with the modeled geometric albedo spectrum of Earth from the NASA Astrobiology Institute's Virtual Planetary Laboratory (VPL) 3-D spectral Earth model \citep{Robinson_2010}; the geometric albedo of Earth in filter A is noticeably higher than that of the model.

\begin{figure}
    \centering
    \includegraphics[width=0.5\linewidth]{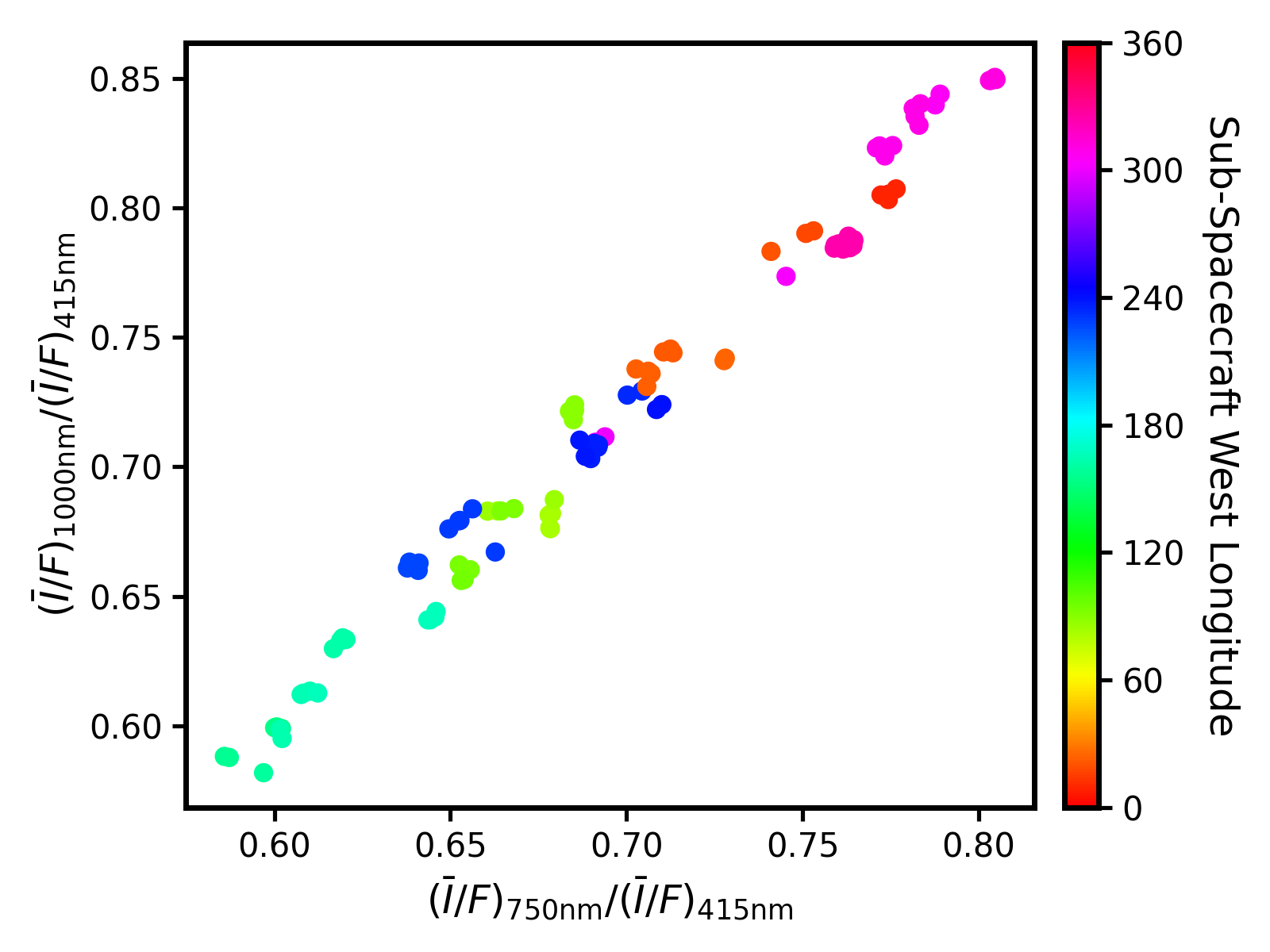}
    \caption{Low-phase angle ($\alpha<45\degree$) measurements of Earth's $\bar{I}/F$ in color-color space. Points are colored by their sub-spacecraft longitude.}
    \label{fig:colorcolor}
\end{figure}

Low-phase angle measurements of Earth in color-color space are shown in Figure~\ref{fig:colorcolor}. We select as our colors the ratio of $\bar{I}/F$'s between filters B and A, and the ratio between filters E and A, because these wavelengths enable comparison with \textit{Galileo} data at similar wavelength ranges \citep{Strauss_2024}. Larger values indicate redder observations.

\begin{figure}[b]
    \centering
    \includegraphics[width=0.5\linewidth]{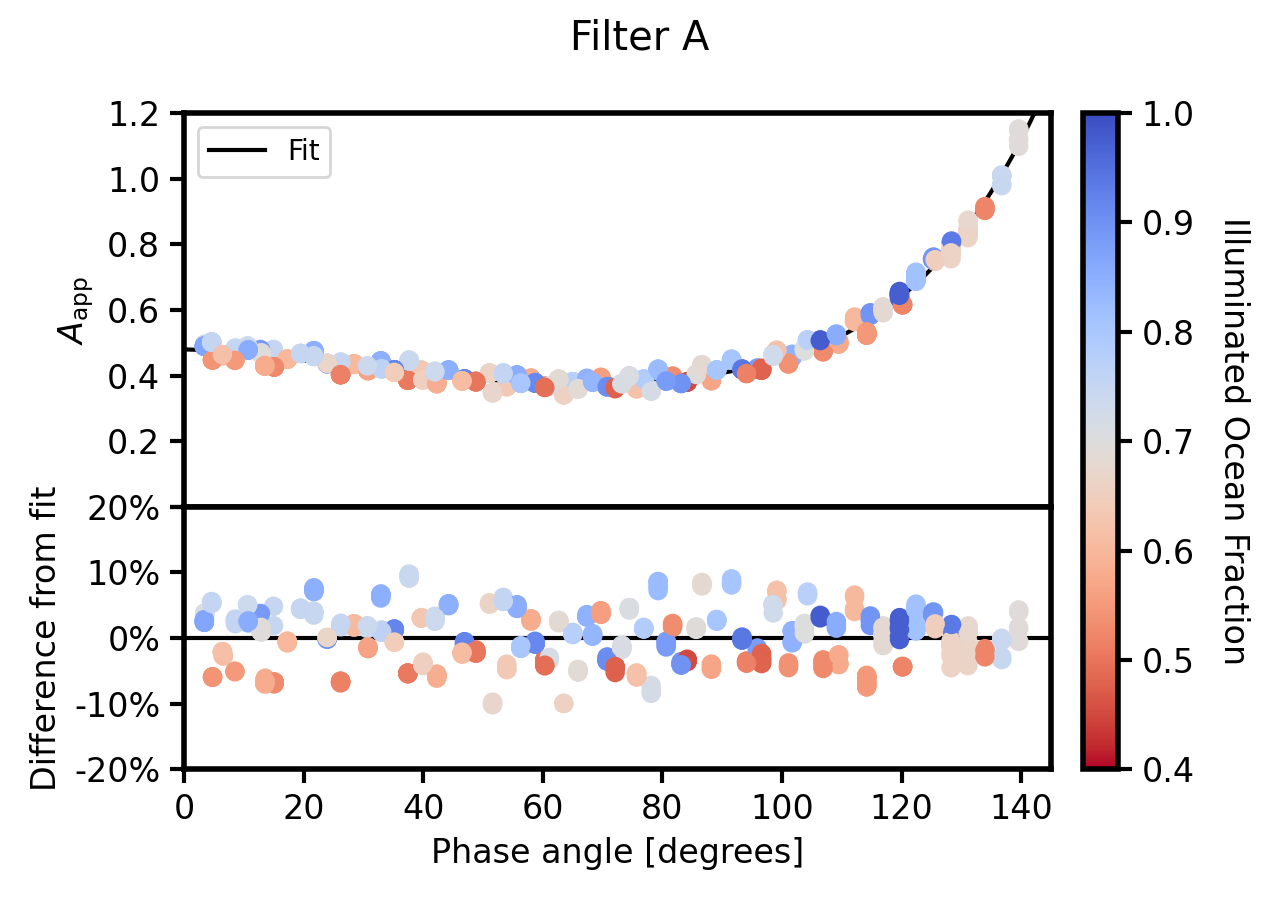}
    \includegraphics[width=0.5\linewidth]{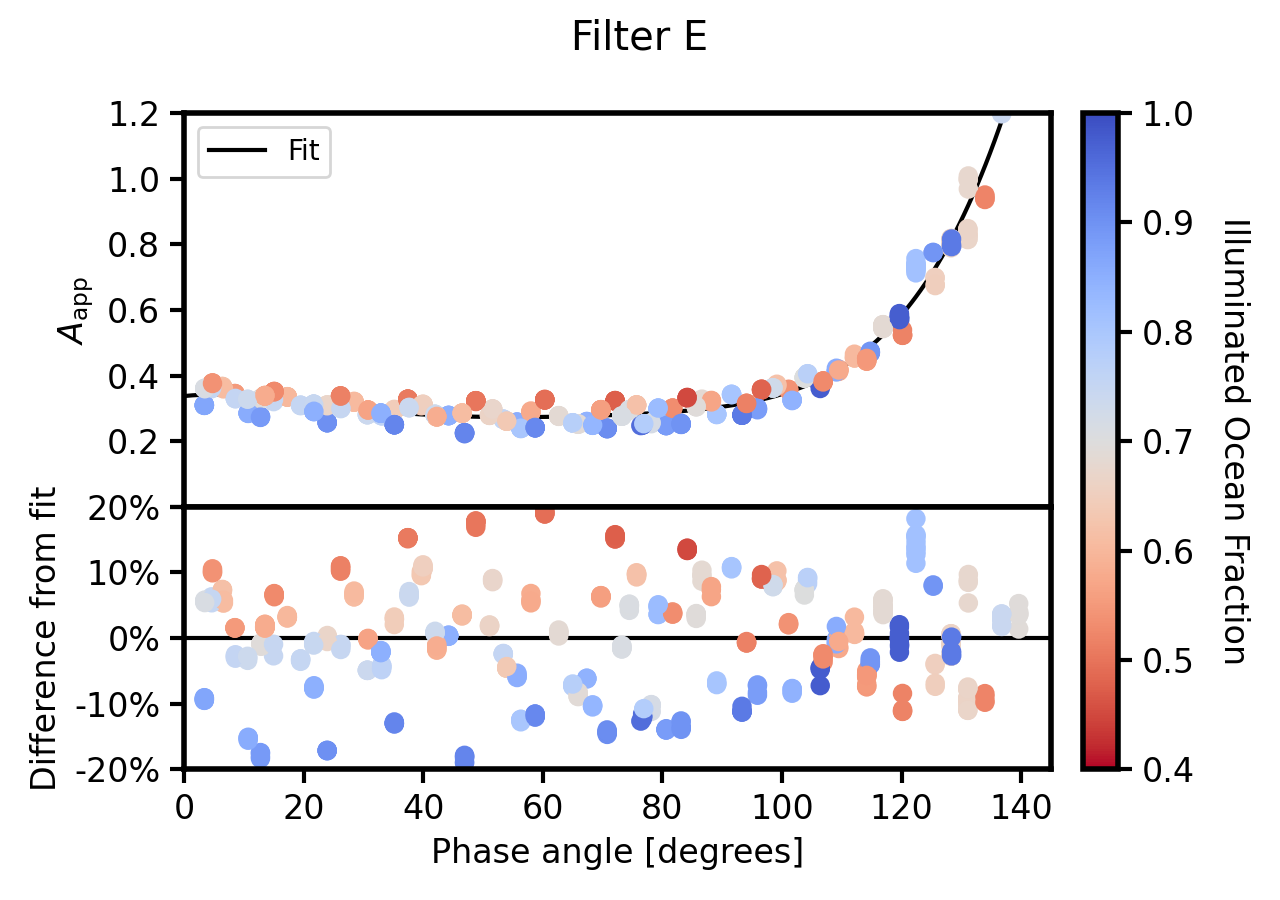}
    \caption{The apparent albedo phase curves of Earth at UV/Vis filters A (top) and E (bottom). Data points are colored based on the fraction of the illuminated disk that is land/ocean. Fitted 5th-degree polynomials are also shown. A sub-plot shows the fractional difference in $A_{\rm{app}}$ between each point and the fitted curve.}
    \label{fig:landfrac}
\end{figure}

In Figure~\ref{fig:landfrac}, we show the apparent albedo phase curves of Earth in filters A and E, as well as the fractional difference between the data and the fitted curve. Points are colored according to their illuminated ocean fractions. We omit corresponding phase curves for filters B, C, and D, as the data are quite similar to those of filter E. Apparent albedo is generally constant with phase angle for phase angles below 100$\degree$. In all phase curves, the apparent albedo varies about the corresponding fit curve by up to $\pm20\%$.

\begin{figure}
    \centering
    \includegraphics[width=0.5\linewidth]{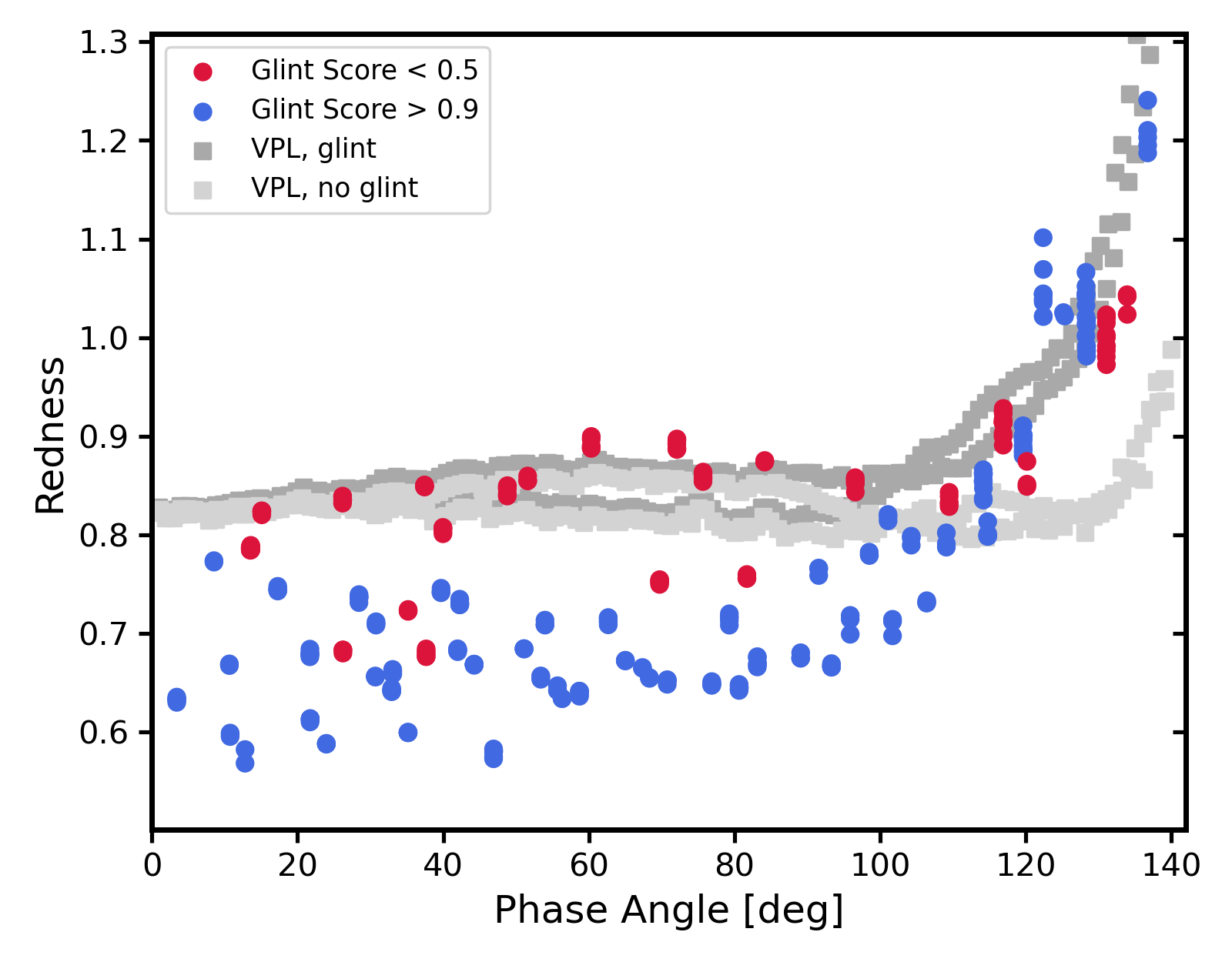}
    \caption{Color phase curve of Earth, where color is defined as the ratio of Earth's $\bar{I}/F$ in filter E to that in filter A. Data are color-coded based on how much of the predicted glint spot falls on the ocean: points where this value exceeds 0.9 are shown in blue; points where this value is less than 0.5 are shown in red; all other measurements are excluded. The extent of the glint spot is calculated using a falloff factor of 0.5 (see Appendix~\hyperref[sec:appendix_a]{A}). Color values derived from the VPL Earth model are also shown, both including (dark gray) and excluding glint (light gray). Values from either model follow two tracks because points are simulated over a full orbit of Earth, so different northern spring and fall seasons manifest different curves.}
    \label{fig:glintcolor}
\end{figure}

In Figure~\ref{fig:glintcolor}, the measured color of Earth is plotted as a function of phase angle. Color in this case is defined as the ratio of the reflectivity in filter E to that in filter A, so redder measurements lie higher on the graph. Included for comparison are the same colors derived by convolving VPL Earth model spectra, both with and without glint, with the \textit{Clementine} filters. At phase angles below $100\degree$, land-dominated observations are in good agreement with either VPL model, while ocean-dominated Earth observations appear bluer than predicted by the model. This effect is also seen in the geometric albedo spectrum in Figure~\ref{fig:geoalb}. The discrepancy between the VPL model and Earth spectral data was discussed in \citet{Robinson_2011}, who used Earth observations from the \textit{EPOXI} mission, and suggested the difference may be due to the scattering of light within the upper layer of the planet's oceans ("volume scattering"), which is not accounted for by the model. At high phase angles, the model predictions diverge, and the Earth measurements, both ocean- and land-dominated, agree with the glinting model. We note that the point at which Earth's redness begins to increase, near $\alpha=100\degree$, is approximately the point in the phase curves in Figure~\ref{fig:phase_curves} where Earth's reflectance begins to be super-Lambertian.

\section{Discussion}\label{sec:discussion}

\subsection{Brightness and Color Variability \label{disc:var}}

Changes in cloud cover, and changes in the illuminated disk land/ocean fractions as the Earth rotates, both contribute to the observed variability in Figure~\ref{fig:landfrac}. The variability in the \textit{Clementine} data over the course of a day was about 10-15\% at phase angles below $90\degree$. This is consistent with the rotational variability of Earth observed in \textit{EPOXI} data \citep{Livengood_2011}, but a direct comparison is difficult because the phase angles in the \textit{Clementine} data changed by up to $10\degree$ over 24 hours. Within each phase curve, variability is approximately constant with phase angle. This contrasts with predictions from previous studies using Earth models, which predict minimal variability of ~5\% near full phase and up to 40\% at crescent phases \citep{Oakley_2009,Robinson_2010}. Our phase-dependent variability measures also contrast with \citet{Robinson_2025}, who found that existing Earthshine data and spacecraft observations of Earth are best fit by a model with increasing variability above phase angles of $\sim110\degree$. It is challenging to draw definitive conclusions about these different observational results as the \textit{Clementine} cubes analyzed here span only 1 month of time, whereas the Earthshine observations incorporated into the analysis of \citet{Robinson_2025} were sourced over multiple years and capture a fuller extent of cloud variability on Earth.

In filter A, we observe the least amount of variability ($\pm10\%$) in Earth's phase curve. This filter has weak surface sensitivity due to strong Rayleigh scattering, and so variability is largely due to varying cloud coverages on the disk, especially higher-altitude, thick clouds. Individual apparent albedo measurements are generally greater in high-ocean fraction observations at all phase angles. This is consistent with our assumption of cloud-driven variability, as cloud fractions are generally higher over oceans than over land \citep{Stubenrauch_2013,King_2013}.

In the remaining filters, we observe variability about the fit curves of $\pm20\%$. While these filters are sensitive to cloud fraction, as in filter A, they are also more sensitive to the surface, and measured variability is also driven by changing disk land/ocean fractions. This includes filter D, despite it overlapping with the broad water vapor absorption band near 950\,nm. High apparent albedo measurements are correlated with high land fractions at phase angles below $100\degree$ (Figure~\ref{fig:landfrac}). Above $120\degree$, we observe a weak correlation between ocean fraction and apparent albedo in each phase curve, though the number of cubes at such phase angles is limited.

The rotational color variability of Earth is well-demonstrated by Figure~\ref{fig:colorcolor}. The data form a line in color-color space, with the redder end occupied by cubes acquired over land-dominated regions (Africa) and the bluer end occupied by ocean-dominated cubes (Pacific Ocean). These color results mirror the trends derived by \citet{Strauss_2024} (their Figure 6). However, we observe a greater range of color values. This may be due to the broader time period of \textit{Clementine} images capturing greater cloud variability than the \textit{Galileo} data, which were acquired over the course of a single day.

\subsection{Glint}\label{subsec:disc_glint}

Previous modeling work \citep{Williams_2008} has predicted that the reflected flux due to glint will be greatest near phase angles of $150\degree$. Additionally, reflectivity due to glint should be greatest at redder wavelengths, as reflected blue light is attenuated by Rayleigh scattering as it passes through the atmosphere \citep{Robinson_2010,Zugger_2011}. Thus, glint may be detectable in \textit{Clementine} disk-averaged measurements by looking for redness correlated with visible ocean fraction in high-phase angle observations. We look for such a trend in the presented \textit{Clementine} dataset.

The \textit{Clementine} dataset contained two sets of observations at extreme phase angles where the measured redness was best matched by the glinting model, despite the majority of the glint spot falling over land (Figure~\ref{fig:glintcolor}: there were 9 sets of observations at $131.1\degree$ where the glint spot fell over northern Australia, with an average glint score of 0.318 and an average redness of 1; and there were 3 sets of observations at $133.9\degree$ where the glint spot fell near the Arabian peninsula, with an average glint score of 0.214 and an average redness of 1.036. Both sets of observations fall over large desert regions, so the observed redness was initially attributed to the presence of these deserts; sands, and especially iron oxide-rich sands, are bright in longwave visible and near-infrared light \citep[e.g.][]{Bullard_2002,Rushby_2020}, and their reflectance can have a significant specular component \citep{Deering_1990}. However, inspections of the resolved observations showed that pixel redness over deserts was less than the disk-integrated color and couldn't explain the whole-disk redness. We suggest the high redness of these observations are due largely to clouds.

\begin{figure}
    \centering
    \includegraphics[width=0.5\linewidth]{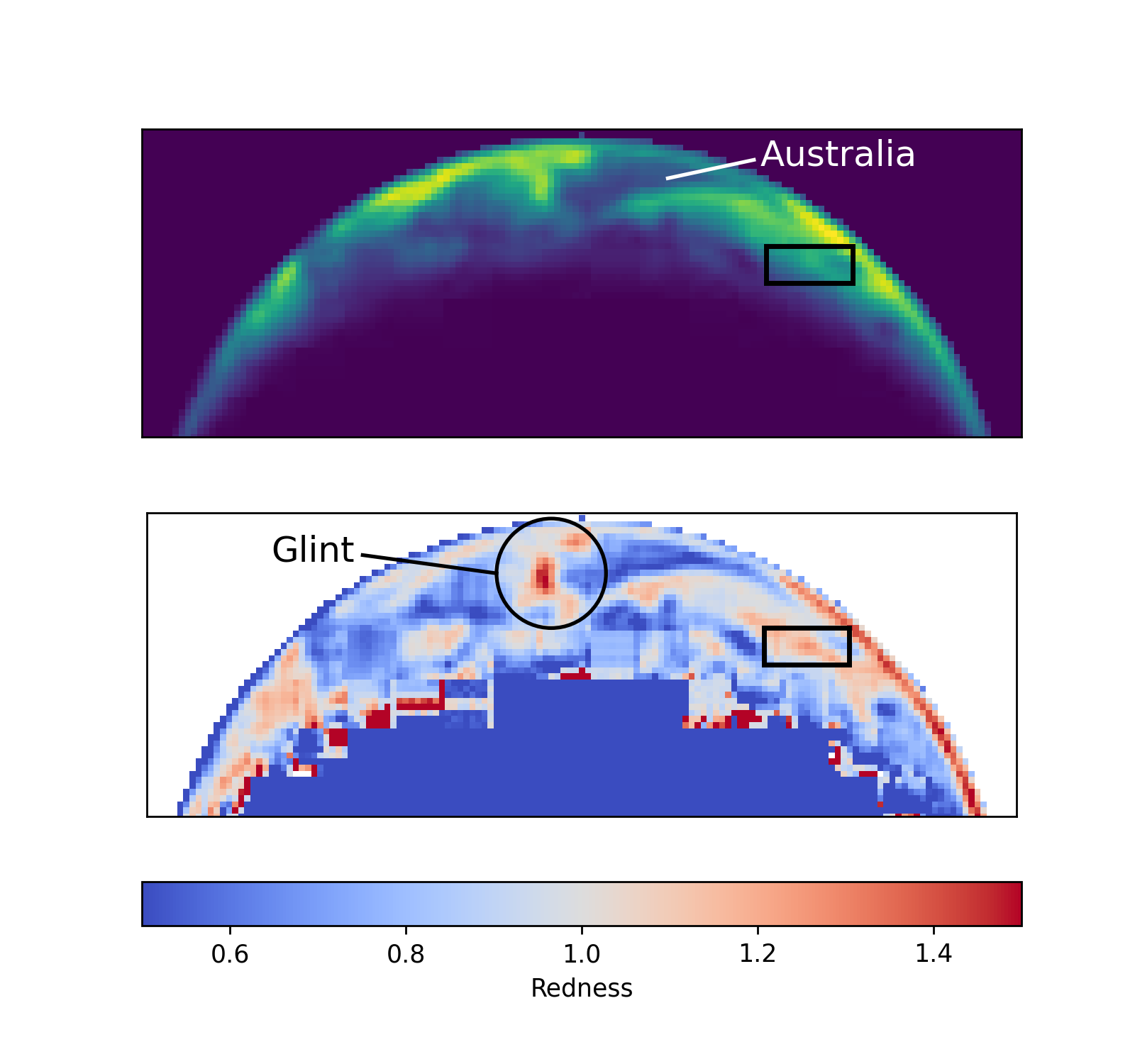}
    \caption{Top: a high-phase \textit{Clementine} Australia observation at filter E. The observation has been cropped to focus on the illuminated portion of the Earth. Bottom: the spatially-resolved redness of the same observation. Rednesses are calculated by dividing pixel-wise the $I/F$ at filter E by that at filter A. The black box shows the location of a suspected high-redness cloud.}
    \label{fig:aus}
\end{figure}

% \begin{figure}
%     \centering
%     \includegraphics[width=0.5\linewidth]{images/redgrid.png}
%     \caption{The rednesses of outgoing radiation streams calculated using \texttt{rfast}. Results for Earth models with glint and no clouds (top) and models with optically thick clouds (bottom) are shown. Streams are organized by solar zenith angle, observer zenith angle, and solar-observer azimuth angle. The approximate geometries of sampled pixels are denoted by black stars.}
%     \label{fig:redgrid}
% \end{figure}

\begin{figure}
    \centering
    \includegraphics[width=0.5\linewidth]{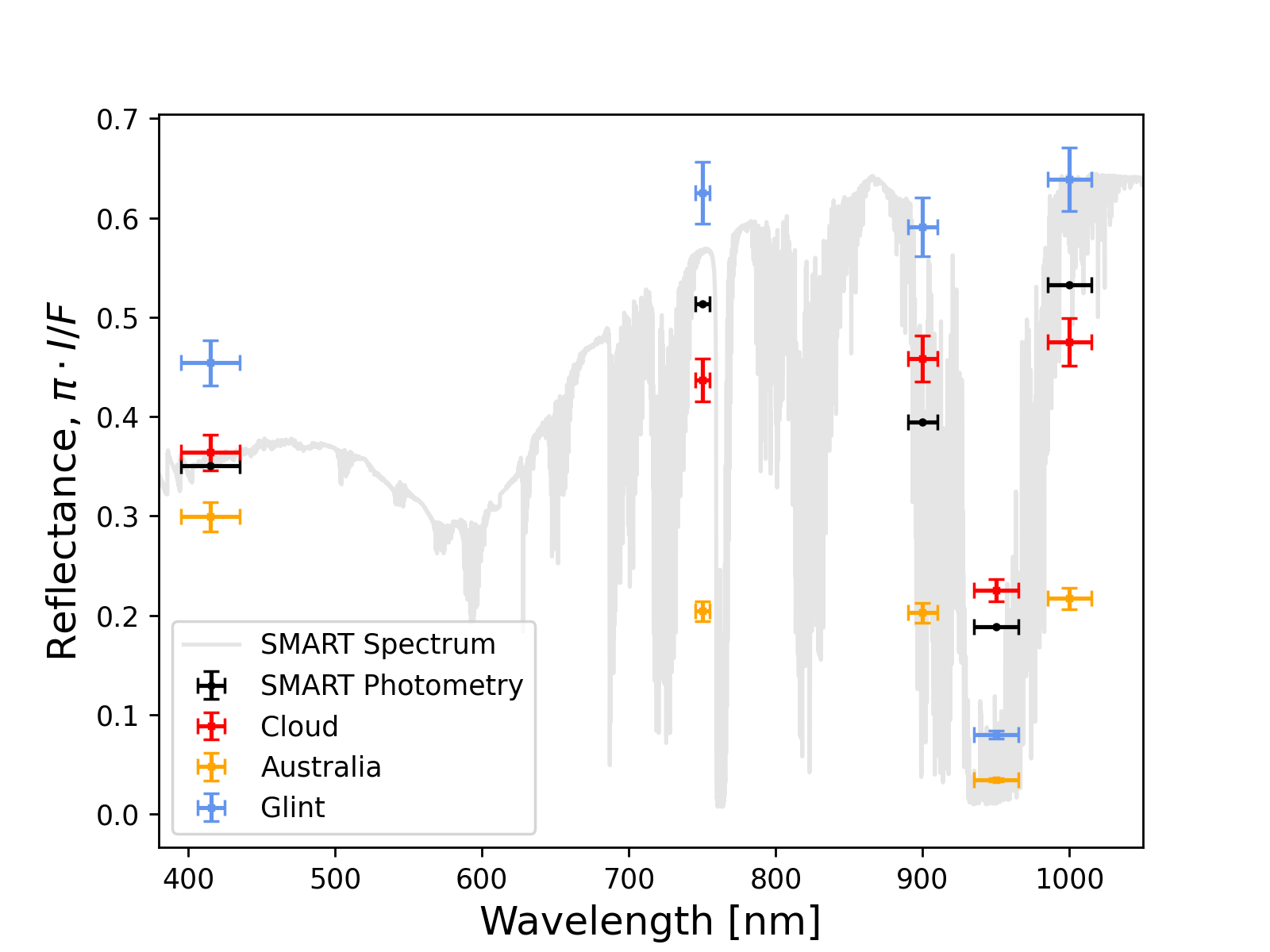}
    \caption{The scaled reflectances of the cloud feature in the previous figure and those produced using \texttt{SMART} convolved with \textit{Clementine} filters A and E. Observed reflectances are shown with 5\% errorbars for the instrument calibration uncertainty. The full modeled spectrum from \texttt{SMART} is also shown in grey. The \texttt{SMART} model was produced using observational geometry similar to that of the boxed region in Figure~\ref{fig:aus_spec}. Also shown are the reflectances from pixels containing either the glint feature or Australia.}
    \label{fig:aus_spec}
\end{figure}

To investigate the potential of clouds to explain the observed redness, we sampled the solar zenith angles, observer zenith angles, and solar-observer azimuths of a small number of pixels which composed suspected cloud features in one Australia observation, shown in Figure~\ref{fig:aus}. We generated reflectance spectra of Earth with optically thick ($\tau=5$) clouds using the spectral forward modeling tool \texttt{SMART} \citep[developed by D.~Crisp;][]{1996JGR...101.4595M}. We took the radiation stream with the geometry that most closely matched that of the outlined area in Figure~\ref{fig:aus}, convolved them with the \textit{Clementine} filters, and compared the model colors to the measured colors (Figure~\ref{fig:aus_spec}). We found that the color of suspected cloud features in the solid box in Figure~\ref{fig:aus} was approximately reproduced by the cloudy Earth scenario.

\subsection{Considerations for Future Work}

% Phase angle limitation
Models have predicted that the contribution from glint to Earth's reflectivity is strongest in observations near a phase angles of $150\degree$. The dataset of \textit{Clementine} cubes were acquired at phase angles up to $140\degree$, including phase angles where glint contributes significantly to the reflectivity of Earth, but they do not access phase angles where glint is predicted to contribute the most. Because of this limitation, we are unable to make broad conclusions about the detectability of glint in monochromatic Earth observations. Future glint studies would greatly benefit from full-disk observations of Earth at phase angles closer to $150\degree$ to better determine the phase and wavelength dependencies of glint in Earth observations.

% Analogue for direct imaging data/ 
%Disk-averaged observations of Earth are a useful testing grounds for interpreting future direct-imaging data from facilities like HWO. Because the data in this work are disk-averaged, they emulate observations of a point-source, which is how future direct imaging of exoplanets will appear. 

% Color for glint detection
Although glint can produce an increased apparent albedo in planetary observations, it is not unique in that regard. Other scatterers\,---\,like clouds or thin aerosols \citep{Robinson_2025}, hazes \citep{GarciaMunoz_2017,Cooper_2025}, or surface ice \citep{Cowan_2009}\,---\,produce degenerate effects. This will complicate the identification of glint in exoplanetary phase curves. Color measurements provide a promising alternate avenue for glint identification due to its distinct redness that increases with phase angle. Indeed, the limited high-phase disk-averaged Earth measurements presented here favor a glinting model over a non-glinting model and suggest that glint reddening effects may be detectable in point source observations of ocean-covered exoplanets. However, interpretations of broadband photometry are challenging, and future high-phase spectral observations of Earth would help to disentangle glint reddening from other scattering effects.

\section{Conclusions}\label{sec:conclusions}

In this work, we present whole-disk observations of Earth at a wide range of geometries acquired with NASA's \textit{Clementine} lunar orbiter. Five spectrally distinct apparent albedo phase curves of Earth are presented. We conducted analyses of these phase curves and measurements of Earth's color to determine if glint is observable in point-source observations of Earth. Our main findings are as follows:

\begin{itemize}
    \item We produced 2,705 phase-dependent measurements of Earth's reflectivity, each at one of five UV/Vis filters, centered on 415\,nm, 750\,nm, 900\,nm, 950\,nm, and 1000\,nm.
    \item We used phase-dependent measurements of Earth to produce apparent albedo phase curves of Earth. These phase curves showcase Earth's non-Lambertian scattering behavior, in particular strong forward scattering at high phase angles ($\alpha>100\degree$).
    \item We approximated Earth's geometric albedo by extrapolating fits to Earth's phase curves to full phase. For UV/Vis filters A through E, we found $A_{\rm g}=0.320,0.223,0.225,0.120,$ and $0.226$, respectively.
    % $A_{\rm g}=0.314,0.222,0.225,0.122,$ and $0.226$
    \item We found the color of Earth in \textit{Clementine} cubes, defined as $(I/F)_{\rm 1000nm}/(I/F)_{\rm 415nm}$, is significantly bluer than the VPL 3-D spectral Earth model at phase angles up to $\alpha\approx120\degree$. Cubes at extreme crescent phases ($\alpha>130\degree$) better agree with the glinting VPL model over the non-glinting model. Cubes with glint may be redder than those without, but this cannot be conclusively shown due to limited data at extreme crescent phases. Additionally, some cubes without glint are as red as those with glint, which may be due to scattering by optically thick clouds.
    \item Earth spectral modeling efforts that incorporate glint would greatly benefit from additional observations of Earth at extreme crescent phases. Earth observations can also be exploited to test and validate methods for characterizing terrestrial exoplanets in advance of future direct imaging missions like HWO.
\end{itemize}

\section{Acknowledgements}\label{sec:ack}

This project was funded by an award from NASA's Exoplanets Research Program (No.\,80NSSC25K7149). \textit{Clementine} data used in this work were acquired from the \href{https://pds-imaging.jpl.nasa.gov/search/}{PDS Image Atlas}. The \texttt{ISIS3} calibration software used in this work is produced by the United States Geological Survey (USGS) and can be found on their \href{https://github.com/DOI-USGS/ISIS3?tab=readme-ov-file}{GitHub page}. No generative AI was used during this work or while preparing this document.

\begin{software}
    \newline astropy \citep{Astropy1,Astropy2,Astropy3}, ISIS3 \citep{Rodriguez2024}, matplotlib \citep{Hunter_2007}, numpy \citep{2020NumPy-Array}, scipy \citep{2020SciPy-NMeth}
\end{software}

\section*{Appendix A: Glint Map Computation}\label{sec:appendix_a}

% Getting better...
Saturated pixel correction requires the production of an accurate glint mask over the \textit{Clementine} image being corrected. Such a mask requires a knowledge of the azimuthal position of the glint spot, relative to the center pixel of a \textit{Clementine} cube, which is not provided by the calibrated files. We present our method for calculating the mask below. 

\renewcommand{\thefigure}{A1}
\begin{figure*}[t!]
    \centering
    \includegraphics[width=0.5\linewidth]{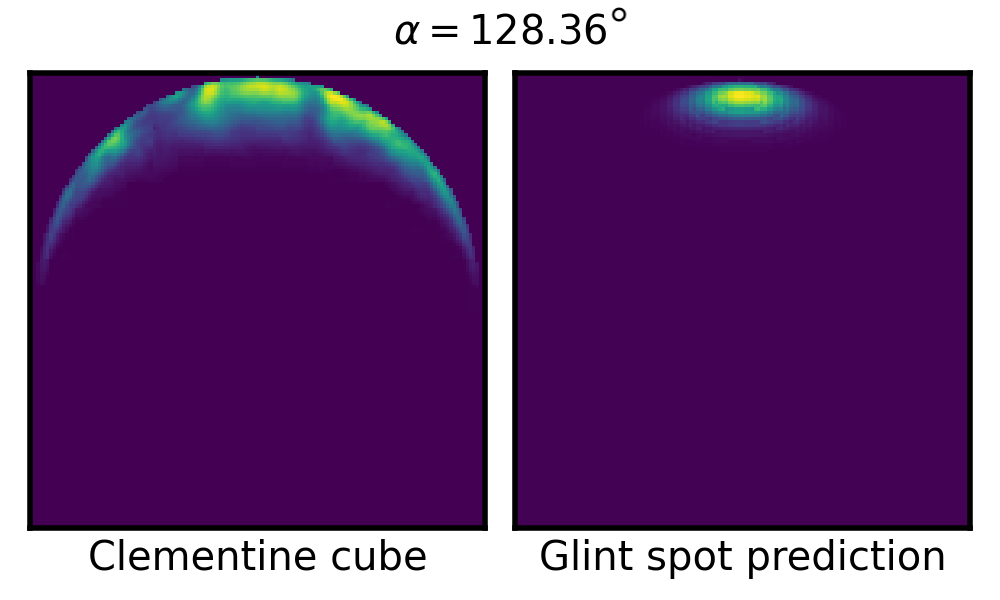}
    \caption{A example of the glint spot prediction for a \textit{Clementine} cube.}
    \label{fig:glintspot}
\end{figure*}

% Glint map calculation
To produce a glint spot mask for a \textit{Clementine} image, we start by creating an empty array that is the same size as the \textit{Clementine} image. Each pixel can be assigned 2D coordinates $i,j$ that describe the pixel's offset from the image center, in units of pixels, in the horizontal and vertical directions. These can be converted to 3D Cartesian coordinates, $(x,y,z)$, on a unit sphere via,
\begin{align*}
    x &= \sqrt{1-i^2-j^2}\\
    y &= i/r\\
    z &= j/r \ ,
\end{align*}
where $r$ is the radius of Earth in pixels. These Cartesian coordinates can then be converted to a longitude-latitude $(\lambda,\theta)$ pair via,
\begin{align*}
    \lambda &= \frac{\pi}{2}-\rm{sign}(x)\cos^{-1}\Bigg(\frac{y}{\sqrt{x^2+y^2}}\Bigg)\\
    \theta &= \sin^{-1}(z) \ .
\end{align*}
The sub-spacecraft point for this glint mask is set to $(\lambda=0,\theta=0)$. Given a sub-solar longitude, and by requiring the sub-solar point lies on the ``illumination equator'', the solar and spacecraft zenith angles ($\mu_{\rm s} and \mu_{\rm o}$) for each grid cell are calculated as the cell's great circle distance (as an angle) from the sub-solar and sub-spacecraft points, respectively. The sun-spacecraft azimuth angle is calculated by converting longitude-latitude pairs to points on a unit sphere as before, then using the treatment in Appendix~\ref{sec:appendix_a} of \citet{Robinson_2025}. These parameters, as well as a wind speed parameter, are used to calculate the ocean bidirectional reflectance distribution function (BRDF) at a given point using the method developed by \citet{Cox_1954}; the I/F from that pixel is,
\begin{align*}
    I/F=\rm BRDF\cdot\cos(\mu_{\rm s})
\end{align*}

The extent of the glint spot is controlled by specifying a ``falloff parameter'' between 0 and 1. Pixels in the array are considered part of the glint spot if the ratio of the calculated $I/F$ to the highest $I/F$ in the map exceeds the falloff parameter. For example, if a falloff parameter of 0.1 is given, the glint spot will contain all pixels with an $I/F$ that is at least 10\% of the largest $I/F$ value.

% Glint mask rotation angle first
The resulting array now holds a predicted glint spot map. We note that the glint spot map necessarily lies on the equator in the frame. Finding the angle of rotation required for fitting the glint spot properly can be accomplished using rotation matrices. Throughout this process, we consider points on the Earth's surface as points on a unit sphere in one of three frames: a \textit{Clementine} frame, where the point with longitude $\lambda=0$ and latitude $\theta=0$ lies at $(1,0,0)$ on the unit sphere and the north pole lies at $(0,0,1)$; a ``north-up'' frame, where the sub-spacecraft point lies at $(1,0,0)$ and points directly to the north lie on the $xz$-plane (this effectively is the frame in which \textit{Clementine} observes points on Earth); and the ``oriented'' frame, where the sub-spacecraft point of a cube is located at $(1,0,0)$ and the sub-solar point lies on the $xy$-plane. Points with longitude $\lambda$ and latitude $\theta$ can be mapped to the unoriented frame by,
\begin{align*}
    x &= \sin(\lambda)\cos(\theta) \\
    y &= \cos(\lambda)\cos(\theta) \\
    z &= \sin(\theta) \ .
\end{align*}

To build a rotation matrix that can map from the unoriented frame to the \textit{Clementine} frame, we start with two rotations that map the sub-spacecraft point with longitude $\lambda_{\rm o}$ and latitude $\theta_{\rm o}$ to the point $(0,1,0)$,
\begin{align*}
    A &= \begin{bmatrix}
            \cos(-\lambda_{\rm o}) & -\sin(-\lambda_{\rm o}) & 0\\
            \sin(-\lambda_{\rm o}) & \cos(-\lambda_{\rm o}) & 0\\
            0 & 0 & 1
        \end{bmatrix} \\
    B &= \begin{bmatrix}
            1 & 0 & 0 \\
            0 & \cos(-\theta_{\rm o}) & -\sin(-\theta_{\rm o}) \\
            0 & \sin(-\theta_{\rm o}) & \cos(-\theta_{\rm o}) 
         \end{bmatrix} \ .\\
\end{align*}
In effect, this rotation will map points on the oriented frame to their position in the orientation of Earth as viewed by \textit{Clementine}, after the disk has been rotated such that north points ``up'' in the image. A final rotation $C$ by an angle $\gamma$ is needed to map a point to the oriented frame,
\begin{align*}
    C = \begin{bmatrix}
        \cos(\gamma) & 0 & \sin(\gamma) \\
        0 & 1 & 0 \\
        -\sin(\gamma) & 0 & \cos(\gamma)
    \end{bmatrix} \ .
\end{align*}
We solve for $\gamma$ numerically such that it minimizes the sub-solar $z$-coordinate post-transformation. The final rotation matrix $CBA$ successfully maps points in the unoriented frame to their position in the oriented frame. More importantly, the angle $\gamma$ is also the angle by which to rotate the glint map so that it corresponds with the glint spot in the \textit{Clementine} image with north oriented ``up.'' By rotating the glint map by the negative of the angle of rotation that orients north ``up,'' the glint mask now lies in the orientation of the original \textit{Clementine} cube.

\section*{Appendix B: \textit{MESSENGER} Data} \label{sec:appendix_b}

In the early stages of this work, we considered incorporating Earth observations acquired during the \textit{MESSENGER} spacecraft's Earth flyby, using the Mercury Dual-Imaging System's Wide-Angle Camera. The data were acquired in two sets: one set of observations was acquired on August 2nd, 2005, at $\alpha=51\degree$, each acquired at one of 12 filters with central wavelengths between $430\,$nm and $1020\,$nm; and another set of observations was acquired on August 3rd, 2005, at phase angles between $98\degree$ and $107\degree$ each at one of three filters between $480\,$nm and $630\,$nm. These data lacked sufficient phase coverage to be applicable to the study of glint in Earth observations, but may be valuable for other applications. The reduced \textit{MESSENGER} data, including disk-averaged $\bar{I}/F$ and apparent albedos, are included in the supplementary data for this work. 

\newpage

\bibliography{biblist}{}

@ARTICLE{Astropy1,
       author = {{Astropy Collaboration} and {Robitaille}, Thomas P. and {Tollerud}, Erik J. and {Greenfield}, Perry and {Droettboom}, Michael and {Bray}, Erik and {Aldcroft}, Tom and {Davis}, Matt and {Ginsburg}, Adam and {Price-Whelan}, Adrian M. and {Kerzendorf}, Wolfgang E. and {Conley}, Alexander and {Crighton}, Neil and {Barbary}, Kyle and {Muna}, Demitri and {Ferguson}, Henry and {Grollier}, Fr{\'e}d{\'e}ric and {Parikh}, Madhura M. and {Nair}, Prasanth H. and {Unther}, Hans M. and {Deil}, Christoph and {Woillez}, Julien and {Conseil}, Simon and {Kramer}, Roban and {Turner}, James E.~H. and {Singer}, Leo and {Fox}, Ryan and {Weaver}, Benjamin A. and {Zabalza}, Victor and {Edwards}, Zachary I. and {Azalee Bostroem}, K. and {Burke}, D.~J. and {Casey}, Andrew R. and {Crawford}, Steven M. and {Dencheva}, Nadia and {Ely}, Justin and {Jenness}, Tim and {Labrie}, Kathleen and {Lim}, Pey Lian and {Pierfederici}, Francesco and {Pontzen}, Andrew and {Ptak}, Andy and {Refsdal}, Brian and {Servillat}, Mathieu and {Streicher}, Ole},
        title = "{Astropy: A community Python package for astronomy}",
      journal = {\aap},
         year = 2013,
        month = oct,
       volume = {558},
          eid = {A33},
        pages = {A33},
          doi = {10.1051/0004-6361/201322068},
archivePrefix = {arXiv},
       eprint = {1307.6212},
 primaryClass = {astro-ph.IM},
       adsurl = {https://ui.adsabs.harvard.edu/abs/2013A&A...558A..33A}
}

@ARTICLE{1996JGR...101.4595M,
       author = {{Meadows}, V.~S. and {Crisp}, D.},
        title = "{Ground-based near-infrared observations of the Venus nightside: The thermal structure and water abundance near the surface}",
      journal = {\jgr},
         year = 1996,
        month = jan,
       volume = {101},
       number = {E2},
        pages = {4595-4622},
          doi = {10.1029/95JE03567},
       adsurl = {https://ui.adsabs.harvard.edu/abs/1996JGR...101.4595M}
}

@ARTICLE{Astropy2,
       author = {{Astropy Collaboration} and {Price-Whelan}, A.~M. and {Sip{\H{o}}cz}, B.~M. and {G{\"u}nther}, H.~M. and {Lim}, P.~L. and {Crawford}, S.~M. and {Conseil}, S. and {Shupe}, D.~L. and {Craig}, M.~W. and {Dencheva}, N. and {Ginsburg}, A. and {VanderPlas}, J.~T. and {Bradley}, L.~D. and {P{\'e}rez-Su{\'a}rez}, D. and {de Val-Borro}, M. and {Aldcroft}, T.~L. and {Cruz}, K.~L. and {Robitaille}, T.~P. and {Tollerud}, E.~J. and {Ardelean}, C. and {Babej}, T. and {Bach}, Y.~P. and {Bachetti}, M. and {Bakanov}, A.~V. and {Bamford}, S.~P. and {Barentsen}, G. and {Barmby}, P. and {Baumbach}, A. and {Berry}, K.~L. and {Biscani}, F. and {Boquien}, M. and {Bostroem}, K.~A. and {Bouma}, L.~G. and {Brammer}, G.~B. and {Bray}, E.~M. and {Breytenbach}, H. and {Buddelmeijer}, H. and {Burke}, D.~J. and {Calderone}, G. and {Cano Rodr{\'\i}guez}, J.~L. and {Cara}, M. and {Cardoso}, J.~V.~M. and {Cheedella}, S. and {Copin}, Y. and {Corrales}, L. and {Crichton}, D. and {D'Avella}, D. and {Deil}, C. and {Depagne}, {\'E}. and {Dietrich}, J.~P. and {Donath}, A. and {Droettboom}, M. and {Earl}, N. and {Erben}, T. and {Fabbro}, S. and {Ferreira}, L.~A. and {Finethy}, T. and {Fox}, R.~T. and {Garrison}, L.~H. and {Gibbons}, S.~L.~J. and {Goldstein}, D.~A. and {Gommers}, R. and {Greco}, J.~P. and {Greenfield}, P. and {Groener}, A.~M. and {Grollier}, F. and {Hagen}, A. and {Hirst}, P. and {Homeier}, D. and {Horton}, A.~J. and {Hosseinzadeh}, G. and {Hu}, L. and {Hunkeler}, J.~S. and {Ivezi{\'c}}, {\v{Z}}. and {Jain}, A. and {Jenness}, T. and {Kanarek}, G. and {Kendrew}, S. and {Kern}, N.~S. and {Kerzendorf}, W.~E. and {Khvalko}, A. and {King}, J. and {Kirkby}, D. and {Kulkarni}, A.~M. and {Kumar}, A. and {Lee}, A. and {Lenz}, D. and {Littlefair}, S.~P. and {Ma}, Z. and {Macleod}, D.~M. and {Mastropietro}, M. and {McCully}, C. and {Montagnac}, S. and {Morris}, B.~M. and {Mueller}, M. and {Mumford}, S.~J. and {Muna}, D. and {Murphy}, N.~A. and {Nelson}, S. and {Nguyen}, G.~H. and {Ninan}, J.~P. and {N{\"o}the}, M. and {Ogaz}, S. and {Oh}, S. and {Parejko}, J.~K. and {Parley}, N. and {Pascual}, S. and {Patil}, R. and {Patil}, A.~A. and {Plunkett}, A.~L. and {Prochaska}, J.~X. and {Rastogi}, T. and {Reddy Janga}, V. and {Sabater}, J. and {Sakurikar}, P. and {Seifert}, M. and {Sherbert}, L.~E. and {Sherwood-Taylor}, H. and {Shih}, A.~Y. and {Sick}, J. and {Silbiger}, M.~T. and {Singanamalla}, S. and {Singer}, L.~P. and {Sladen}, P.~H. and {Sooley}, K.~A. and {Sornarajah}, S. and {Streicher}, O. and {Teuben}, P. and {Thomas}, S.~W. and {Tremblay}, G.~R. and {Turner}, J.~E.~H. and {Terr{\'o}n}, V. and {van Kerkwijk}, M.~H. and {de la Vega}, A. and {Watkins}, L.~L. and {Weaver}, B.~A. and {Whitmore}, J.~B. and {Woillez}, J. and {Zabalza}, V. and {Astropy Contributors}},
        title = "{The Astropy Project: Building an Open-science Project and Status of the v2.0 Core Package}",
      journal = {\aj},
         year = 2018,
        month = sep,
       volume = {156},
       number = {3},
          eid = {123},
        pages = {123},
          doi = {10.3847/1538-3881/aabc4f},
archivePrefix = {arXiv},
       eprint = {1801.02634},
 primaryClass = {astro-ph.IM},
       adsurl = {https://ui.adsabs.harvard.edu/abs/2018AJ....156..123A}
}

@ARTICLE{Astropy3,
       author = {{Astropy Collaboration} and {Price-Whelan}, Adrian M. and {Lim}, Pey Lian and {Earl}, Nicholas and {Starkman}, Nathaniel and {Bradley}, Larry and {Shupe}, David L. and {Patil}, Aarya A. and {Corrales}, Lia and {Brasseur}, C.~E. and {N{\"o}the}, Maximilian and {Donath}, Axel and {Tollerud}, Erik and {Morris}, Brett M. and {Ginsburg}, Adam and {Vaher}, Eero and {Weaver}, Benjamin A. and {Tocknell}, James and {Jamieson}, William and {van Kerkwijk}, Marten H. and {Robitaille}, Thomas P. and {Merry}, Bruce and {Bachetti}, Matteo and {G{\"u}nther}, H. Moritz and {Aldcroft}, Thomas L. and {Alvarado-Montes}, Jaime A. and {Archibald}, Anne M. and {B{\'o}di}, Attila and {Bapat}, Shreyas and {Barentsen}, Geert and {Baz{\'a}n}, Juanjo and {Biswas}, Manish and {Boquien}, M{\'e}d{\'e}ric and {Burke}, D.~J. and {Cara}, Daria and {Cara}, Mihai and {Conroy}, Kyle E. and {Conseil}, Simon and {Craig}, Matthew W. and {Cross}, Robert M. and {Cruz}, Kelle L. and {D'Eugenio}, Francesco and {Dencheva}, Nadia and {Devillepoix}, Hadrien A.~R. and {Dietrich}, J{\"o}rg P. and {Eigenbrot}, Arthur Davis and {Erben}, Thomas and {Ferreira}, Leonardo and {Foreman-Mackey}, Daniel and {Fox}, Ryan and {Freij}, Nabil and {Garg}, Suyog and {Geda}, Robel and {Glattly}, Lauren and {Gondhalekar}, Yash and {Gordon}, Karl D. and {Grant}, David and {Greenfield}, Perry and {Groener}, Austen M. and {Guest}, Steve and {Gurovich}, Sebastian and {Handberg}, Rasmus and {Hart}, Akeem and {Hatfield-Dodds}, Zac and {Homeier}, Derek and {Hosseinzadeh}, Griffin and {Jenness}, Tim and {Jones}, Craig K. and {Joseph}, Prajwel and {Kalmbach}, J. Bryce and {Karamehmetoglu}, Emir and {Ka{\l}uszy{\'n}ski}, Miko{\l}aj and {Kelley}, Michael S.~P. and {Kern}, Nicholas and {Kerzendorf}, Wolfgang E. and {Koch}, Eric W. and {Kulumani}, Shankar and {Lee}, Antony and {Ly}, Chun and {Ma}, Zhiyuan and {MacBride}, Conor and {Maljaars}, Jakob M. and {Muna}, Demitri and {Murphy}, N.~A. and {Norman}, Henrik and {O'Steen}, Richard and {Oman}, Kyle A. and {Pacifici}, Camilla and {Pascual}, Sergio and {Pascual-Granado}, J. and {Patil}, Rohit R. and {Perren}, Gabriel I. and {Pickering}, Timothy E. and {Rastogi}, Tanuj and {Roulston}, Benjamin R. and {Ryan}, Daniel F. and {Rykoff}, Eli S. and {Sabater}, Jose and {Sakurikar}, Parikshit and {Salgado}, Jes{\'u}s and {Sanghi}, Aniket and {Saunders}, Nicholas and {Savchenko}, Volodymyr and {Schwardt}, Ludwig and {Seifert-Eckert}, Michael and {Shih}, Albert Y. and {Jain}, Anany Shrey and {Shukla}, Gyanendra and {Sick}, Jonathan and {Simpson}, Chris and {Singanamalla}, Sudheesh and {Singer}, Leo P. and {Singhal}, Jaladh and {Sinha}, Manodeep and {Sip{\H{o}}cz}, Brigitta M. and {Spitler}, Lee R. and {Stansby}, David and {Streicher}, Ole and {{\v{S}}umak}, Jani and {Swinbank}, John D. and {Taranu}, Dan S. and {Tewary}, Nikita and {Tremblay}, Grant R. and {de Val-Borro}, Miguel and {Van Kooten}, Samuel J. and {Vasovi{\'c}}, Zlatan and {Verma}, Shresth and {de Miranda Cardoso}, Jos{\'e} Vin{\'\i}cius and {Williams}, Peter K.~G. and {Wilson}, Tom J. and {Winkel}, Benjamin and {Wood-Vasey}, W.~M. and {Xue}, Rui and {Yoachim}, Peter and {Zhang}, Chen and {Zonca}, Andrea and {Astropy Project Contributors}},
        title = "{The Astropy Project: Sustaining and Growing a Community-oriented Open-source Project and the Latest Major Release (v5.0) of the Core Package}",
      journal = {\apj},
         year = 2022,
        month = aug,
       volume = {935},
       number = {2},
          eid = {167},
        pages = {167},
          doi = {10.3847/1538-4357/ac7c74},
archivePrefix = {arXiv},
       eprint = {2206.14220},
 primaryClass = {astro-ph.IM},
       adsurl = {https://ui.adsabs.harvard.edu/abs/2022ApJ...935..167A}
}

@ARTICLE{2020SciPy-NMeth,
  author  = {Virtanen, Pauli and Gommers, Ralf and Oliphant, Travis E. and
            Haberland, Matt and Reddy, Tyler and Cournapeau, David and
            Burovski, Evgeni and Peterson, Pearu and Weckesser, Warren and
            Bright, Jonathan and {van der Walt}, St{\'e}fan J. and
            Brett, Matthew and Wilson, Joshua and Millman, K. Jarrod and
            Mayorov, Nikolay and Nelson, Andrew R. J. and Jones, Eric and
            Kern, Robert and Larson, Eric and Carey, C J and
            Polat, {\.I}lhan and Feng, Yu and Moore, Eric W. and
            {VanderPlas}, Jake and Laxalde, Denis and Perktold, Josef and
            Cimrman, Robert and Henriksen, Ian and Quintero, E. A. and
            Harris, Charles R. and Archibald, Anne M. and
            Ribeiro, Ant{\^o}nio H. and Pedregosa, Fabian and
            {van Mulbregt}, Paul and {SciPy 1.0 Contributors}},
  title   = {{{SciPy} 1.0: Fundamental Algorithms for Scientific
            Computing in Python}},
  journal = {Nature Methods},
  year    = {2020},
  volume  = {17},
  pages   = {261--272},
  adsurl  = {https://rdcu.be/b08Wh},
  doi     = {10.1038/s41592-019-0686-2},
}

@ARTICLE{2020NumPy-Array,
  author  = {Harris, Charles R. and Millman, K. Jarrod and
            van der Walt, Stéfan J and Gommers, Ralf and
            Virtanen, Pauli and Cournapeau, David and
            Wieser, Eric and Taylor, Julian and Berg, Sebastian and
            Smith, Nathaniel J. and Kern, Robert and Picus, Matti and
            Hoyer, Stephan and van Kerkwijk, Marten H. and
            Brett, Matthew and Haldane, Allan and
            Fernández del Río, Jaime and Wiebe, Mark and
            Peterson, Pearu and Gérard-Marchant, Pierre and
            Sheppard, Kevin and Reddy, Tyler and Weckesser, Warren and
            Abbasi, Hameer and Gohlke, Christoph and
            Oliphant, Travis E.},
  title   = {Array programming with {NumPy}},
  journal = {Nature},
  year    = {2020},
  volume  = {585},
  pages   = {357–362},
  doi     = {10.1038/s41586-020-2649-2}
}

@ARTICLE{Hunter_2007,
  author={Hunter, John D.},
  journal={Computing in Science \& Engineering}, 
  title={Matplotlib: A 2D Graphics Environment}, 
  year={2007},
  volume={9},
  number={3},
  pages={90-95},
  doi={10.1109/MCSE.2007.55}
}

@MISC{Rodriguez2024,
  title     = "Integrated software for imagers and spectrometers ({ISIS}) 8.3.0",
  author    = "Rodriguez, Kelvin and {Astrogeology Science Center}",
  publisher = "U.S. Geological Survey",
  year      =  2024
}

@article{ACTON199665,
    title = {Ancillary data services of NASA's Navigation and Ancillary Information Facility},
    journal = {Planetary and Space Science},
    volume = {44},
    number = {1},
    pages = {65-70},
    year = {1996},
    note = {Planetary data system},
    issn = {0032-0633},
    doi = {https://doi.org/10.1016/0032-0633(95)00107-7},
    url = {https://www.sciencedirect.com/science/article/pii/0032063395001077},
    author = {Charles H. Acton}
}

@article{ACTON20189,
    title = {A look towards the future in the handling of space science mission geometry},
    journal = {Planetary and Space Science},
    volume = {150},
    pages = {9-12},
    year = {2018},
    note = {Enabling Open and Interoperable Access to Planetary Science and Heliophysics Databases and Tools},
    issn = {0032-0633},
    doi = {https://doi.org/10.1016/j.pss.2017.02.013},
    url = {https://www.sciencedirect.com/science/article/pii/S0032063316303129},
    author = {Charles Acton and Nathaniel Bachman and Boris Semenov and Edward Wright}
}

@ARTICLE{Hillier1999,
       author = {{Hillier}, John K. and {Buratti}, Bonnie J. and {Hill}, Kathryn},
        title = "{Multispectral Photometry of the Moon and Absolute Calibration of the Clementine UV/Vis Camera}",
      journal = {\icarus},
         year = 1999,
        month = oct,
       volume = {141},
       number = {2},
        pages = {205-225},
          doi = {10.1006/icar.1999.6184},
       adsurl = {https://ui.adsabs.harvard.edu/abs/1999Icar..141..205H}
}

@ARTICLE{Shkuratov2001,
       author = {{Shkuratov}, Yu. G. and {Kaidash}, V.~G. and {Kreslavsky}, M.~A. and {Opanasenko}, N.~V.},
        title = "{Absolute Calibration of the Clementine UVVIS Data: Comparison with Ground-Based Observation of the Moon}",
      journal = {Solar System Research},
         year = 2001,
        month = jan,
       volume = {35},
       number = {1},
        pages = {29-34},
       adsurl = {https://ui.adsabs.harvard.edu/abs/2001SoSyR..35...29S}
}

@ARTICLE{Ohtake2010,
       author = {{Ohtake}, Makiko and {Matsunaga}, Tsuneo and {Yokota}, Yasuhiro and {Yamamoto}, Satoru and {Ogawa}, Yoshiko and {Morota}, Tomokatsu and {Honda}, Chikatoshi and {Haruyama}, Junichi and {Kitazato}, Kouhei and {Takeda}, Hiroshi and {Iwasaki}, Akira and {Nakamura}, Ryousuke and {Hiroi}, Takahiro and {Kodama}, Sinsuke and {Otake}, Hisashi},
        title = "{Deriving the Absolute Reflectance of Lunar Surface Using SELENE (Kaguya) Multiband Imager Data}",
      journal = {\ssr},
         year = 2010,
        month = jul,
       volume = {154},
       number = {1-4},
        pages = {57-77},
          doi = {10.1007/s11214-010-9689-0},
       adsurl = {https://ui.adsabs.harvard.edu/abs/2010SSRv..154...57O}
}

@article{Strauss_2024,
    doi = {10.3847/1538-3881/ad1bd1},
    url = {https://dx.doi.org/10.3847/1538-3881/ad1bd1},
    year = {2024},
    month = {feb},
    publisher = {The American Astronomical Society},
    volume = {167},
    number = {3},
    pages = {87},
    author = {Ryder H. Strauss and Tyler D. Robinson and David E. Trilling and Ryan Cummings and Christopher J. Smith},
    title = {Exoplanet Analog Observations of Earth from Galileo Disk-integrated Photometry},
    journal = {The Astronomical Journal}
}

@article{Canny_1986,
    author = {Canny, John},
    year = {1986},
    month = {12},
    pages = {679 - 698},
    title = {A Computational Approach To Edge Detection},
    volume = {PAMI-8},
    isbn = {9780080515816},
    journal = {Pattern Analysis and Machine Intelligence, IEEE Transactions on},
    doi = {10.1109/TPAMI.1986.4767851}
}

@inproceedings{xie_2002,
  title={A new efficient ellipse detection method},
  author={Xie, Yonghong and Ji, Qiang},
  booktitle={2002 International Conference on Pattern Recognition},
  volume={2},
  pages={957--960},
  year={2002},
  organization={IEEE}
}

@ARTICLE{Russell1916,
       author = {{Russell}, Henry Norris},
        title = "{On the Albedo of the Planets and Their Satellites}",
      journal = {\apj},
         year = 1916,
        month = apr,
       volume = {43},
        pages = {173-196},
          doi = {10.1086/142244},
       adsurl = {https://ui.adsabs.harvard.edu/abs/1916ApJ....43..173R}
}

@ARTICLE{Robinson_2010,
       author = {{Robinson}, Tyler D. and {Meadows}, Victoria S. and {Crisp}, David},
        title = "{Detecting Oceans on Extrasolar Planets Using the Glint Effect}",
      journal = {\apjl},
         year = 2010,
        month = sep,
       volume = {721},
       number = {1},
        pages = {L67-L71},
          doi = {10.1088/2041-8205/721/1/L67},
archivePrefix = {arXiv},
       eprint = {1008.3864},
 primaryClass = {astro-ph.EP},
       adsurl = {https://ui.adsabs.harvard.edu/abs/2010ApJ...721L..67R}
}

@ARTICLE{Williams_2008,
       author = {{Williams}, Darren M. and {Gaidos}, Eric},
        title = "{Detecting the glint of starlight on the oceans of distant planets}",
      journal = {\icarus},
         year = 2008,
        month = jun,
       volume = {195},
       number = {2},
        pages = {927-937},
          doi = {10.1016/j.icarus.2008.01.002},
archivePrefix = {arXiv},
       eprint = {0801.1852},
 primaryClass = {astro-ph},
       adsurl = {https://ui.adsabs.harvard.edu/abs/2008Icar..195..927W}
}

@article {Arking_1968,
      author = "Albert  {Arking} and John  {Potter}",
      title = "The Phase Curve of Venus and the Nature of its Clouds",
      journal = "Journal of Atmospheric Sciences",
      year = "1968",
      publisher = "American Meteorological Society",
      address = "Boston MA, USA",
      volume = "25",
      number = "4",
      doi = "10.1175/1520-0469(1968)025<0617:TPCOVA>2.0.CO;2",
      pages=      "617 - 628",
      url = "https://journals.ametsoc.org/view/journals/atsc/25/4/1520-0469_1968_025_0617_tpcova_2_0_co_2.xml"
}

@INPROCEEDINGS{Kordas_1995,
       author = {{Kordas}, Joseph F. and {Lewis}, Isabella T. and {Priest}, Robert E. and {White}, W. Travis and {Nielsen}, Darron P. and {Park}, Hye-Sook and {Wilson}, Bruce A. and {Shannon}, Michael J. and {Ledebuhr}, Arno G. and {Pleasance}, Lyn D.},
        title = "{UV/visible camera for the Clementine mission}",
    booktitle = {Space Telescopes and Instruments},
         year = 1995,
       editor = {{Bely}, Pierre Y. and {Breckinridge}, James B.},
       series = {Society of Photo-Optical Instrumentation Engineers (SPIE) Conference Series},
       volume = {2478},
        month = jun,
        pages = {175-186},
          doi = {10.1117/12.210943},
       adsurl = {https://ui.adsabs.harvard.edu/abs/1995SPIE.2478..175K}
}

@ARTICLE{Cooper_2025,
       author = {{Cooper}, Chase and {Robinson}, Tyler D. and {Barnes}, Jason W. and {Mayorga}, L.~C. and {Robinthal}, Lily},
        title = "{Extreme Forward Scattering Observed in Disk-averaged Near-infrared Phase Curves of Titan}",
      journal = {\psj},
         year = 2025,
        month = oct,
       volume = {6},
       number = {10},
          eid = {228},
        pages = {228},
          doi = {10.3847/PSJ/ae071f},
archivePrefix = {arXiv},
       eprint = {2507.00924},
 primaryClass = {astro-ph.EP},
       adsurl = {https://ui.adsabs.harvard.edu/abs/2025PSJ.....6..228C}
}

@ARTICLE{GarciaMunoz_2017,
       author = {{Garc{\'\i}a Mu{\~n}oz}, A. and {Lavvas}, P. and {West}, R.~A.},
        title = "{Titan brighter at twilight than in daylight}",
      journal = {Nature Astronomy},
         year = 2017,
        month = apr,
       volume = {1},
          eid = {0114},
        pages = {0114},
          doi = {10.1038/s41550-017-0114},
archivePrefix = {arXiv},
       eprint = {1704.07460},
 primaryClass = {astro-ph.EP},
       adsurl = {https://ui.adsabs.harvard.edu/abs/2017NatAs...1E.114G}
}

@ARTICLE{Robinson_2025,
       author = {{Robinson}, Tyler D.},
        title = "{Inferring and Interpreting the Visual Geometric Albedo and Phase Function of Earth}",
      journal = {\psj},
         year = 2026,
        month = jan,
       volume = {7},
       number = {1},
          eid = {12},
        pages = {12},
          doi = {10.3847/PSJ/ae2ec6},
archivePrefix = {arXiv},
       eprint = {2507.22258},
 primaryClass = {astro-ph.EP},
       adsurl = {https://ui.adsabs.harvard.edu/abs/2026PSJ.....7...12R}
}

@ARTICLE{Zugger_2010,
       author = {{Zugger}, M.~E. and {Kasting}, J.~F. and {Williams}, D.~M. and {Kane}, T.~J. and {Philbrick}, C.~R.},
        title = "{Light Scattering from Exoplanet Oceans and Atmospheres}",
      journal = {\apj},
         year = 2010,
        month = nov,
       volume = {723},
       number = {2},
        pages = {1168-1179},
          doi = {10.1088/0004-637X/723/2/1168},
archivePrefix = {arXiv},
       eprint = {1006.3525},
 primaryClass = {astro-ph.EP},
       adsurl = {https://ui.adsabs.harvard.edu/abs/2010ApJ...723.1168Z}
}

@ARTICLE{Zugger_2011,
       author = {{Zugger}, M.~E. and {Kasting}, J.~F. and {Williams}, D.~M. and {Kane}, T.~J. and {Philbrick}, C.~R.},
        title = "{Searching for Water Earths in the Near-infrared}",
      journal = {\apj},
         year = 2011,
        month = sep,
       volume = {739},
       number = {1},
          eid = {12},
        pages = {12},
          doi = {10.1088/0004-637X/739/1/12},
       adsurl = {https://ui.adsabs.harvard.edu/abs/2011ApJ...739...12Z}
}

@ARTICLE{Robinson_2014,
       author = {{Robinson}, Tyler D. and {Ennico}, Kimberly and {Meadows}, Victoria S. and {Sparks}, William and {Bussey}, D. Ben J. and {Schwieterman}, Edward W. and {Breiner}, Jonathan},
        title = "{Detection of Ocean Glint and Ozone Absorption Using LCROSS Earth Observations}",
      journal = {\apj},
         year = 2014,
        month = jun,
       volume = {787},
       number = {2},
          eid = {171},
        pages = {171},
          doi = {10.1088/0004-637X/787/2/171},
archivePrefix = {arXiv},
       eprint = {1405.4557},
 primaryClass = {astro-ph.EP},
       adsurl = {https://ui.adsabs.harvard.edu/abs/2014ApJ...787..171R}
}

@inproceedings{feinbergetal2024,
    author = {Lee Feinberg and John Ziemer and Megan Ansdell and Julie Crooke and Courtney Dressing and Bertrand Mennesson and John O'Meara and Joshua Pepper and Aki Roberge},
    title = {{The Habitable Worlds Observatory engineering view: status, plans, and opportunities}},
    volume = {13092},
    booktitle = {Space Telescopes and Instrumentation 2024: Optical, Infrared, and Millimeter Wave},
    editor = {Laura E. Coyle and Shuji Matsuura and Marshall D. Perrin},
    organization = {International Society for Optics and Photonics},
    publisher = {SPIE},
    pages = {130921N},
    year = {2024},
    doi = {10.1117/12.3018328},
    URL = {https://doi.org/10.1117/12.3018328}
}

@ARTICLE{Kempton_2023,
       author = {{Kempton}, Eliza M.-R. and {Zhang}, Michael and {Bean}, Jacob L. and {Steinrueck}, Maria E. and {Piette}, Anjali A.~A. and {Parmentier}, Vivien and {Malsky}, Isaac and {Roman}, Michael T. and {Rauscher}, Emily and {Gao}, Peter and et al.},
        title = "{A reflective, metal-rich atmosphere for GJ 1214b from its JWST phase curve}",
      journal = {\nat},
         year = 2023,
        month = aug,
       volume = {620},
       number = {7972},
        pages = {67-71},
          doi = {10.1038/s41586-023-06159-5},
archivePrefix = {arXiv},
       eprint = {2305.06240},
 primaryClass = {astro-ph.EP},
       adsurl = {https://ui.adsabs.harvard.edu/abs/2023Natur.620...67K}
}

@ARTICLE{Livengood_2011,
       author = {{Livengood}, Timothy A. and {Deming}, L. Drake and {A'Hearn}, Michael F. and {Charbonneau}, David and {Hewagama}, Tilak and {Lisse}, Carey M. and {McFadden}, Lucy A. and {Meadows}, Victoria S. and {Robinson}, Tyler D. and {Seager}, Sara and et al.},
        title = "{Properties of an Earth-Like Planet Orbiting a Sun-Like Star: Earth Observed by the EPOXI Mission}",
      journal = {Astrobiology},
         year = 2011,
        month = nov,
       volume = {11},
       number = {9},
        pages = {907-930},
          doi = {10.1089/ast.2011.0614},
       adsurl = {https://ui.adsabs.harvard.edu/abs/2011AsBio..11..907L}
}

@ARTICLE{Cox_1954,
       author = {{Cox}, Charles and {Munk}, Walter},
        title = "{Measurement of the Roughness of the Sea Surface from Photographs of the Sun's Glitter}",
      journal = {Journal of the Optical Society of America (1917-1983)},
         year = 1954,
        month = nov,
       volume = {44},
       number = {11},
        pages = {838},
          doi = {10.1364/JOSA.44.000838},
       adsurl = {https://ui.adsabs.harvard.edu/abs/1954JOSA...44..838C}
}

@ARTICLE{King_2013,
       author = {{King}, Michael D. and {Platnick}, Steven and {Menzel}, W. Paul and {Ackerman}, Steven A. and {Hubanks}, Paul A.},
        title = "{Spatial and Temporal Distribution of Clouds Observed by MODIS Onboard the Terra and Aqua Satellites}",
      journal = {IEEE Transactions on Geoscience and Remote Sensing},
         year = 2013,
        month = jul,
       volume = {51},
       number = {7},
        pages = {3826-3852},
          doi = {10.1109/TGRS.2012.2227333},
       adsurl = {https://ui.adsabs.harvard.edu/abs/2013ITGRS..51.3826K}
}

@ARTICLE{Cowan_2009,
       author = {{Cowan}, Nicolas B. and {Agol}, Eric and {Meadows}, Victoria S. and {Robinson}, Tyler and {Livengood}, Timothy A. and {Deming}, Drake and {Lisse}, Carey M. and {A'Hearn}, Michael F. and {Wellnitz}, Dennis D. and {Seager}, Sara and {Charbonneau}, David and {EPOXI Team}},
        title = "{Alien Maps of an Ocean-bearing World}",
      journal = {\apj},
         year = 2009,
        month = aug,
       volume = {700},
       number = {2},
        pages = {915-923},
          doi = {10.1088/0004-637X/700/2/915},
archivePrefix = {arXiv},
       eprint = {0905.3742},
 primaryClass = {astro-ph.EP},
       adsurl = {https://ui.adsabs.harvard.edu/abs/2009ApJ...700..915C}
}

@article{Vaughn_2023,
    author = {Vaughan, Sophia R and Gebhard, Timothy D and Bott, Kimberly and Casewell, Sarah L and Cowan, Nicolas B and Doelman, David S and Kenworthy, Matthew and Mazoyer, Johan and Millar-Blanchaer, Maxwell A and Trees, Victor J H and Stam, Daphne M and Absil, Olivier and Altinier, Lisa and Baudoz, Pierre and Belikov, Ruslan and Bidot, Alexis and Birkby, Jayne L and Bonse, Markus J and Brandl, Bernhard and Carlotti, Alexis and Choquet, Elodie and van Dam, Dirk and Desai, Niyati and Fogarty, Kevin and Fowler, J and van Gorkom, Kyle and Gutierrez, Yann and Guyon, Olivier and Haffert, Sebastiaan Y and Herscovici-Schiller, Olivier and Hours, Adrien and Juanola-Parramon, Roser and Kleisioti, Evangelia and König, Lorenzo and van Kooten, Maaike and Krasteva, Mariya and Laginja, Iva and Landman, Rico and Leboulleux, Lucie and Mouillet, David and N’Diaye, Mamadou and Por, Emiel H and Pueyo, Laurent and Snik, Frans},
    title = {Chasing rainbows and ocean glints: Inner working angle constraints for the Habitable Worlds Observatory},
    journal = {Monthly Notices of the Royal Astronomical Society},
    volume = {524},
    number = {4},
    pages = {5477-5485},
    year = {2023},
    month = {08},
    issn = {0035-8711},
    doi = {10.1093/mnras/stad2127},
    url = {https://doi.org/10.1093/mnras/stad2127},
    eprint = {https://academic.oup.com/mnras/article-pdf/524/4/5477/51031631/stad2127.pdf},
}

@ARTICLE{Robinson_2011,
       author = {{Robinson}, Tyler D. and {Meadows}, Victoria S. and {Crisp}, David and {Deming}, Drake and {A'Hearn}, Michael F. and {Charbonneau}, David and {Livengood}, Timothy A. and {Seager}, Sara and {Barry}, Richard K. and {Hearty}, Thomas and {Hewagama}, Tilak and {Lisse}, Carey M. and {McFadden}, Lucy A. and {Wellnitz}, Dennis D.},
        title = "{Earth as an Extrasolar Planet: Earth Model Validation Using EPOXI Earth Observations}",
      journal = {Astrobiology},
         year = 2011,
        month = jun,
       volume = {11},
       number = {5},
        pages = {393-408},
          doi = {10.1089/ast.2011.0642},
       adsurl = {https://ui.adsabs.harvard.edu/abs/2011AsBio..11..393R}
}

@ARTICLE{Danjon_1936,
       author = {{Danjon}, Andre},
        title = "{Nouvelles recherches sur la photometrie de la lumiere cendree et l' albedo de la terre}",
      journal = {Annalen der Kaiserlichen Universitats-Sternwarte in Strassburg},
         year = 1936,
        month = jan,
       volume = {3},
        pages = {139-179},
       adsurl = {https://ui.adsabs.harvard.edu/abs/1936AnStr...3..139D}
}

@ARTICLE{Palle_2003,
       author = {{Pall{\'e}}, E. and {Goode}, P.~R. and {Yurchyshyn}, V. and {Qiu}, J. and {Hickey}, J. and {Monta{\~n}{\'e}S Rodriguez}, P. and {Chu}, M.-C. and {Kolbe}, E. and {Brown}, C.~T. and {Koonin}, S.~E.},
        title = "{Earthshine and the Earth's albedo: 2. Observations and simulations over 3 years}",
      journal = {Journal of Geophysical Research (Atmospheres)},
         year = 2003,
        month = nov,
       volume = {108},
       number = {D22},
          eid = {4710},
        pages = {4710},
          doi = {10.1029/2003JD003611},
       adsurl = {https://ui.adsabs.harvard.edu/abs/2003JGRD..108.4710P}
}

@ARTICLE{Goode_2001,
       author = {{Goode}, P.~R. and {Qiu}, J. and {Yurchyshyn}, V. and {Hickey}, J. and {Chu}, M.-C. and {Kolbe}, E. and {Brown}, C.~T. and {Koonin}, S.~E.},
        title = "{Earthshine observations of the Earth's reflectance}",
      journal = {\grl},
         year = 2001,
        month = may,
       volume = {28},
       number = {9},
        pages = {1671-1674},
          doi = {10.1029/2000GL012580},
       adsurl = {https://ui.adsabs.harvard.edu/abs/2001GeoRL..28.1671G}
}

@article{Qiu_2003,
author = {Qiu, J. and Goode, P. R. and Pallé, E. and Yurchyshyn, V. and Hickey, J. and Montañés Rodriguez, P. and Chu, M.-C. and Kolbe, E. and Brown, C. T. and Koonin, S. E.},
title = {Earthshine and the Earth's albedo: 1. Earthshine observations and measurements of the lunar phase function for accurate measurements of the Earth's Bond albedo},
journal = {Journal of Geophysical Research: Atmospheres},
volume = {108},
number = {D22},
pages = {},
doi = {https://doi.org/10.1029/2003JD003610},
url = {https://agupubs.onlinelibrary.wiley.com/doi/abs/10.1029/2003JD003610},
eprint = {https://agupubs.onlinelibrary.wiley.com/doi/pdf/10.1029/2003JD003610},
year = {2003}
}

@ARTICLE{Stubenrauch_2013,
       author = {{Stubenrauch}, C.~J. and {Rossow}, W.~B. and {Kinne}, S. and {Ackerman}, S. and {Cesana}, G. and {Chepfer}, H. and {Di Girolamo}, L. and {Getzewich}, B. and {Guignard}, A. and {Heidinger}, A. and et al.},
        title = "{Assessment of Global Cloud Datasets from Satellites: Project and Database Initiated by the GEWEX Radiation Panel}",
      journal = {Bulletin of the American Meteorological Society},
         year = 2013,
        month = jul,
       volume = {94},
       number = {7},
        pages = {1031-1049},
          doi = {10.1175/BAMS-D-12-00117.1},
       adsurl = {https://ui.adsabs.harvard.edu/abs/2013BAMS...94.1031S}
}

@ARTICLE{Oakley_2009,
       author = {{Oakley}, P.~H.~H. and {Cash}, W.},
        title = "{Construction of an Earth Model: Analysis of Exoplanet Light Curves and Mapping the Next Earth with the New Worlds Observer}",
      journal = {\apj},
         year = 2009,
        month = aug,
       volume = {700},
       number = {2},
        pages = {1428-1439},
          doi = {10.1088/0004-637X/700/2/1428},
       adsurl = {https://ui.adsabs.harvard.edu/abs/2009ApJ...700.1428O}
}

@article{Young_1973,
title = {Are the clouds of venus sulfuric acid?},
journal = {Icarus},
volume = {18},
number = {4},
pages = {564-582},
year = {1973},
issn = {0019-1035},
doi = {https://doi.org/10.1016/0019-1035(73)90059-6},
url = {https://www.sciencedirect.com/science/article/pii/0019103573900596},
author = {A.T. Young}
}

@ARTICLE{Coffeen_1969,
       author = {{Coffeen}, D.~L. and {Gehrels}, T.},
        title = "{Wavelength Dependence of Polarization. XV. Observations of Venus}",
      journal = {\aj},
         year = 1969,
        month = apr,
       volume = {74},
        pages = {433},
          doi = {10.1086/110822},
       adsurl = {https://ui.adsabs.harvard.edu/abs/1969AJ.....74..433C}
}

@ARTICLE{Franklin_1965,
       author = {{Franklin}, F.~A. and {Cook}, F.~A.},
        title = "{Optical properties of Saturn's rings. II. Two-color phase curves of the two bright rings.}",
      journal = {\aj},
         year = 1965,
        month = nov,
       volume = {70},
        pages = {704},
          doi = {10.1086/109806},
       adsurl = {https://ui.adsabs.harvard.edu/abs/1965AJ.....70..704F}
}

@ARTICLE{Showalter_1992,
       author = {{Showalter}, Mark R. and {Pollack}, James B. and {Ockert}, Maureen E. and {Doyle}, Laurance R. and {Dalton}, J.~B.},
        title = "{A photometric study of Saturn's Ring}",
      journal = {\icarus},
         year = 1992,
        month = dec,
       volume = {100},
       number = {2},
        pages = {394-411},
          doi = {10.1016/0019-1035(92)90107-I},
       adsurl = {https://ui.adsabs.harvard.edu/abs/1992Icar..100..394S}
}

@ARTICLE{Poulet_2002,
       author = {{Poulet}, F. and {Cuzzi}, J.~N. and {French}, R.~G. and {Dones}, L.},
        title = "{A Study of Saturn's Ring Phase Curves from HST Observations}",
      journal = {\icarus},
         year = 2002,
        month = jul,
       volume = {158},
       number = {1},
        pages = {224-248},
          doi = {10.1006/icar.2002.6852},
       adsurl = {https://ui.adsabs.harvard.edu/abs/2002Icar..158..224P}
}

@ARTICLE{Rushby_2020,
       author = {{Rushby}, Andrew J. and {Shields}, Aomawa L. and {Wolf}, Eric T. and {Lagu{\"e}}, Marysa and {Burgasser}, Adam},
        title = "{The Effect of Land Albedo on the Climate of Land-dominated Planets in the TRAPPIST-1 System}",
      journal = {\apj},
         year = 2020,
        month = dec,
       volume = {904},
       number = {2},
          eid = {124},
        pages = {124},
          doi = {10.3847/1538-4357/abbe04},
archivePrefix = {arXiv},
       eprint = {2011.03621},
 primaryClass = {astro-ph.EP},
       adsurl = {https://ui.adsabs.harvard.edu/abs/2020ApJ...904..124R}
}

@ARTICLE{Bullard_2002,
       author = {{Bullard}, Joanna E. and {White}, Kevin},
        title = "{Quantifying iron oxide coatings on dune sands using spectrometric measurements: An example from the Simpson-Strzelecki Desert, Australia}",
      journal = {Journal of Geophysical Research (Solid Earth)},
         year = 2002,
        month = jun,
       volume = {107},
       number = {B6},
          eid = {2125},
        pages = {2125},
          doi = {10.1029/2001JB000454},
       adsurl = {https://ui.adsabs.harvard.edu/abs/2002JGRB..107.2125B}
}

@ARTICLE{Wong_2020,
       author = {{Wong}, Ian and {Shporer}, Avi and {Kitzmann}, Daniel and {Morris}, Brett M. and {Heng}, Kevin and {Hoeijmakers}, H. Jens and {Demory}, Brice-Olivier and {Ahlers}, John P. and {Mansfield}, Megan and {Bean}, Jacob L. and et al.},
        title = "{Exploring the Atmospheric Dynamics of the Extreme Ultrahot Jupiter KELT-9b Using TESS Photometry}",
      journal = {\aj},
         year = 2020,
        month = aug,
       volume = {160},
       number = {2},
          eid = {88},
        pages = {88},
          doi = {10.3847/1538-3881/aba2cb},
archivePrefix = {arXiv},
       eprint = {1910.01607},
 primaryClass = {astro-ph.EP},
       adsurl = {https://ui.adsabs.harvard.edu/abs/2020AJ....160...88W}
}

@ARTICLE{Mikal-Evans_2023,
       author = {{Mikal-Evans}, Thomas and {Sing}, David K. and {Dong}, Jiayin and {Foreman-Mackey}, Daniel and {Kataria}, Tiffany and {Barstow}, Joanna K. and {Goyal}, Jayesh M. and {Lewis}, Nikole K. and {Lothringer}, Joshua D. and {Mayne}, Nathan J. and et al.},
        title = "{A JWST NIRSpec Phase Curve for WASP-121b: Dayside Emission Strongest Eastward of the Substellar Point and Nightside Conditions Conducive to Cloud Formation}",
      journal = {\apjl},
         year = 2023,
        month = feb,
       volume = {943},
       number = {2},
          eid = {L17},
        pages = {L17},
          doi = {10.3847/2041-8213/acb049},
archivePrefix = {arXiv},
       eprint = {2301.03209},
 primaryClass = {astro-ph.EP},
       adsurl = {https://ui.adsabs.harvard.edu/abs/2023ApJ...943L..17M}
}

@ARTICLE{Knutson_2012,
       author = {{Knutson}, Heather A. and {Lewis}, Nikole and {Fortney}, Jonathan J. and {Burrows}, Adam and {Showman}, Adam P. and {Cowan}, Nicolas B. and {Agol}, Eric and {Aigrain}, Suzanne and {Charbonneau}, David and {Deming}, Drake and et al.},
        title = "{3.6 and 4.5 {\ensuremath{\mu}}m Phase Curves and Evidence for Non-equilibrium Chemistry in the Atmosphere of Extrasolar Planet HD 189733b}",
      journal = {\apj},
         year = 2012,
        month = jul,
       volume = {754},
       number = {1},
          eid = {22},
        pages = {22},
          doi = {10.1088/0004-637X/754/1/22},
archivePrefix = {arXiv},
       eprint = {1206.6887},
 primaryClass = {astro-ph.EP},
       adsurl = {https://ui.adsabs.harvard.edu/abs/2012ApJ...754...22K}
}

@ARTICLE{Dang_2025,
       author = {{Dang}, Lisa and {Bell}, Taylor J. and {Shu}, Ying (Zoe) and {Cowan}, Nicolas B. and {Bean}, Jacob L. and {Deming}, Drake and {Kempton}, Eliza M.-R. and {Mansfield}, Megan Weiner and {Rauscher}, Emily and {Parmentier}, Vivien and et al.},
        title = "{A Comprehensive Analysis of Spitzer 4.5 {\ensuremath{\mu}}m Phase Curves of Hot Jupiters}",
      journal = {\aj},
         year = 2025,
        month = jan,
       volume = {169},
       number = {1},
          eid = {32},
        pages = {32},
          doi = {10.3847/1538-3881/ad8dd7},
archivePrefix = {arXiv},
       eprint = {2408.13308},
 primaryClass = {astro-ph.EP},
       adsurl = {https://ui.adsabs.harvard.edu/abs/2025AJ....169...32D}
}

@ARTICLE{McCullough_2006,
       author = {{McCullough}, P.~R.},
        title = "{Models of Polarized Light from Oceans and Atmospheres of Earth-like Extrasolar Planets}",
      journal = {arXiv e-prints},
         year = 2006,
        month = oct,
          eid = {astro-ph/0610518},
        pages = {astro-ph/0610518},
          doi = {10.48550/arXiv.astro-ph/0610518},
archivePrefix = {arXiv},
       eprint = {astro-ph/0610518},
 primaryClass = {astro-ph},
       adsurl = {https://ui.adsabs.harvard.edu/abs/2006astro.ph.10518M}
}

@INCOLLECTION{Robinson_2018,
       author = {{Robinson}, Tyler D.},
        title = "{Characterizing Exoplanet Habitability}",
    booktitle = {Handbook of Exoplanets},
         year = 2018,
       editor = {{Deeg}, Hans J. and {Belmonte}, Juan Antonio},
          eid = {67},
        pages = {67},
    publisher = {Springer Nature},
          doi = {10.1007/978-3-319-55333-7_67},
       adsurl = {https://ui.adsabs.harvard.edu/abs/2018haex.bookE..67R}
}

@ARTICLE{Nozette_1994,
       author = {{Nozette}, Stewart and {Rustan}, P. and {Pleasance}, L.~P. and {Horan}, D.~M. and {Regeon}, P. and {Shoemaker}, E.~M. and {Spudis}, P.~D. and {Acton}, C.~H. and {Baker}, D.~N. and {Blamont}, J.~E. and et al.},
        title = "{The Clementine Mission to the Moon: Scientific Overview}",
      journal = {Science},
         year = 1994,
        month = dec,
       volume = {266},
       number = {5192},
        pages = {1835-1839},
          doi = {10.1126/science.266.5192.1835},
       adsurl = {https://ui.adsabs.harvard.edu/abs/1994Sci...266.1835N}
}

@ARTICLE{Turnbull_2006,
       author = {{Turnbull}, Margaret C. and {Traub}, Wesley A. and {Jucks}, Kenneth W. and {Woolf}, Neville J. and {Meyer}, Michael R. and {Gorlova}, Nadya and {Skrutskie}, Michael F. and {Wilson}, John C.},
        title = "{Spectrum of a Habitable World: Earthshine in the Near-Infrared}",
      journal = {\apj},
         year = 2006,
        month = jun,
       volume = {644},
       number = {1},
        pages = {551-559},
          doi = {10.1086/503322},
       adsurl = {https://ui.adsabs.harvard.edu/abs/2006ApJ...644..551T}
}

@ARTICLE{Roccetti_2025c,
       author = {{Roccetti}, Giulia and {Sterzik}, Michael F. and {Emde}, Claudia and {Manev}, Mihail and {Bagnulo}, Stefano and {Seidel}, Julia V.},
        title = "{Planet Earth in reflected and polarized light: III. Modeling and analysis of a decade-long catalog of Earthshine observations}",
      journal = {\aap},
         year = 2025,
        month = oct,
       volume = {702},
          eid = {A262},
        pages = {A262},
          doi = {10.1051/0004-6361/202555758},
archivePrefix = {arXiv},
       eprint = {2509.13415},
 primaryClass = {astro-ph.EP},
       adsurl = {https://ui.adsabs.harvard.edu/abs/2025A&A...702A.262R}
}

@ARTICLE{Deering_1990,
       author = {{Deering}, D.~W. and {Eck}, T.~F. and {Otterman}, J.},
        title = "{Bidirectional reflectances of selected desert surfaces and their three-parameter soil characterization}",
      journal = {Agricultural and Forest Meteorology},
         year = 1990,
        month = jan,
       volume = {52},
       number = {1},
        pages = {71-93},
          doi = {10.1016/0168-1923(90)90101-B},
       adsurl = {https://ui.adsabs.harvard.edu/abs/1990AgFM...52...71D}
}
\bibliographystyle{aasjournal}

\end{document}